\documentclass[aps,prd,longbibliography,twocolumn,preprintnumbers,nofootinbib,
superscriptaddress,amsmath,floatfix]{revtex4}

\usepackage{amssymb}  
\usepackage{graphicx,subfigure}
\usepackage{exscale}
\usepackage{textcomp}
\usepackage{enumerate}
\usepackage [english]{babel}
\usepackage[utf8]{inputenc}
\usepackage [autostyle, english = american]{csquotes}
\MakeOuterQuote{"}

\usepackage{color}

\usepackage[normalem]{ulem}  

\newcommand{\non}{\nonumber\\}

\newcommand{\be}{\begin{equation}}
\newcommand{\ee}{\end{equation}}
\newcommand{\bea}{\begin{eqnarray}}
\newcommand{\eea}{\end{eqnarray}}
\newcommand{\ba}[1]{\begin{array}{#1}}
\newcommand{\ea}{\end{array}}

\newcommand{\Tr}{{\rm Tr}}

\newcommand{\bm}[1]{\mbox{\boldmath${#1}$}}

\renewcommand{\vec}[1]{\bm{#1}}

\begin{document}

\title{Chiral pasta}

\author{Andreas Schmitt}
\email{a.schmitt@soton.ac.uk}
\affiliation{Mathematical Sciences and STAG Research Centre, University of Southampton, Southampton SO17 1BJ, United Kingdom}

\date{30 March 2020}

\begin{abstract} 
Interiors of neutron stars are ultra-dense and may contain a core of deconfined quark matter. Such a core  
connects to the outer layers smoothly or through a sharp microscopic interface or through an intermediate macroscopic layer of inhomogeneous mixed phases, which is globally neutral but locally contains electrically charged domains. Here I employ a nucleon-meson model under neutron star conditions that shows a first-order 
chiral phase transition at large densities. In the vicinity of this chiral  transition I calculate the free energies of various mixed phases -- different 'pasta structures' -- in the Wigner-Seitz approximation. Crucially, chirally broken nuclear matter and the approximately chirally symmetric phase (loosely interpreted as quark matter) are treated on the same footing. This allows me to 
compute the interface profiles of bubbles, rods, and slabs fully consistently, taking into account electromagnetic screening effects and without needing the surface tension as an input. I find that the full numerical results tend to disfavor mixed phases compared to a simple step-like approximation used frequently in the literature and that the predominantly favored pasta structure consists of slabs with a surface tension $\Sigma \simeq 6\, {\rm MeV}/{\rm fm}^2$. 

\end{abstract}

\maketitle


\section{Introduction}

Neutron stars probe a large range of baryon densities, from sub-saturation densities in the outer layers up to several times nuclear saturation density in the core. Quantum Chromodynamics (QCD) predicts the existence of deconfined quark matter at ultra-high densities, but it is unknown from first principles at which density the transition from nuclear matter to quark matter occurs. While theoretically extremely challenging, input to this question can be obtained from 
astrophysical observations. Quark matter has different thermodynamic and transport properties compared to nuclear matter, and thus a sizable quark matter 
core may have observable consequences for masses, radii, cooling behavior, etc., of single neutron stars as well as for gravitational wave signals from neutron 
star mergers.

In this paper, I am interested not primarily in a bulk phase of quark matter, but in the interface that separates nuclear matter from a potential quark matter core. This interface itself, depending on its size and structure, might influence observable properties of the neutron star. Since a continuity between 
quark and hadronic matter is conceivable
\cite{Schafer:1998ef,Hatsuda:2006ps,Schmitt:2010pf,Baym:2019iky,Fujimoto:2019sxg}, there may not be an interface at all but rather a smooth transition. The other possibility, predicted by various phenomenological models, is a first-order phase transition. In this case, there can either be a sharp interface, i.e., the transition from nuclear matter to quark matter occurs via a domain wall of microscopic thickness. Depending on the extent of the discontinuity at the transition, it can induce qualitative changes in the mass-radius curve of the star \cite{Alford:2013aca} and possibly in the signals from neutron star mergers \cite{Most:2018eaw,Han:2018mtj}. Or, the transition can be made less abrupt 
by a macroscopic region of a mixed phase, i.e., an inhomogeneous, possibly crystalline, phase, where quark and hadronic phases are separated spatially and occupy different volume fractions as the density varies \cite{Glendenning:1992vb,glendenningbook,PhysRevLett.70.1355}. Here, quark and hadronic phases each carry nonzero electric charge, such that their overall charge is zero in order to maintain global charge neutrality. The quark-hadron mixed 
phase can affect observable properties of the star through its transport properties \cite{Na:2012td} or through its rigidity, which may result in a sustained ellipticity of a rotating 
neutron star and thus in the emission of gravitational waves \cite{JohnsonMcDaniel:2011rt,JohnsonMcDaniel:2012wg}. A similar 
mixed phase is expected to occur in the crust-core transition region of the star, where a crystalline coexistence of nuclei and a neutron (super)fluid is predicted to turn into homogeneous nuclear matter via different geometric structures as one moves to higher densities \cite{Ravenhall:1983uh,1984PThPh..71..320H}. These structures, for example bubbles, rods, or slabs of one phase immersed in the other, are often referred to as 'nuclear pasta' due to their resemblance with gnocchi, spaghetti, and lasagne \cite{2013NatPh...9..396N,Caplan:2016uvu,Schmitt:2017efp,Pearson:2020bxz}. Here I will study pasta phases at the quark-hadron transition, more precisely at the chiral phase transition\footnote{To keep the terminology simple, I use 'pasta' collectively, including spherical structures. This is slightly different from the use of 'nuclear pasta', which usually does not include the spherical nuclei in the outer crust of a neutron star.}.  

Various levels of approximations have been employed in the literature for mixed phases at the quark-hadron transition. Usually, nuclear and quark matter are treated with two different models \cite{Alford:2001zr,Alford:2002rj,Maruyama:2006jk,Maruyama:2007ey,Bhattacharyya:2009fg,Yasutake:2014oxa,Wu:2017xaz,Maslov:2018ghi,Xia:2019pnq,Weber:2019xvv}.
In this approach, the profile of a quark-hadron interface cannot be calculated 
microscopically since this requires knowledge of an effective potential that connects the two phases. Instead, one may approximate the free energies of different mixed phase geometries by assuming spatially constant profiles on either side of the interface, with the interface itself being step-like \cite{glendenningbook}, and treating the surface tension as an external parameter, either varying this parameter freely or employing constraints from microscopic calculations \cite{Palhares:2010be,Palhares:2011jd,PhysRevD.84.036011,Pinto:2012aq,Mintz:2012mz,Ke:2013wga,Gao:2016hks,Fraga:2018cvr}. 
One can improve on this simple estimate by including the charge screening effect  \cite{Heiselberg:1993dc,Voskresensky:2001jq,Voskresensky:2002hu,Alford:2006bx,Alford:2008ge}. This renders the profiles non-constant, but, if two different models are being used, still requires an external choice for the surface tension. 

\begin{figure} [t]
\begin{center}
\includegraphics[width=\columnwidth]{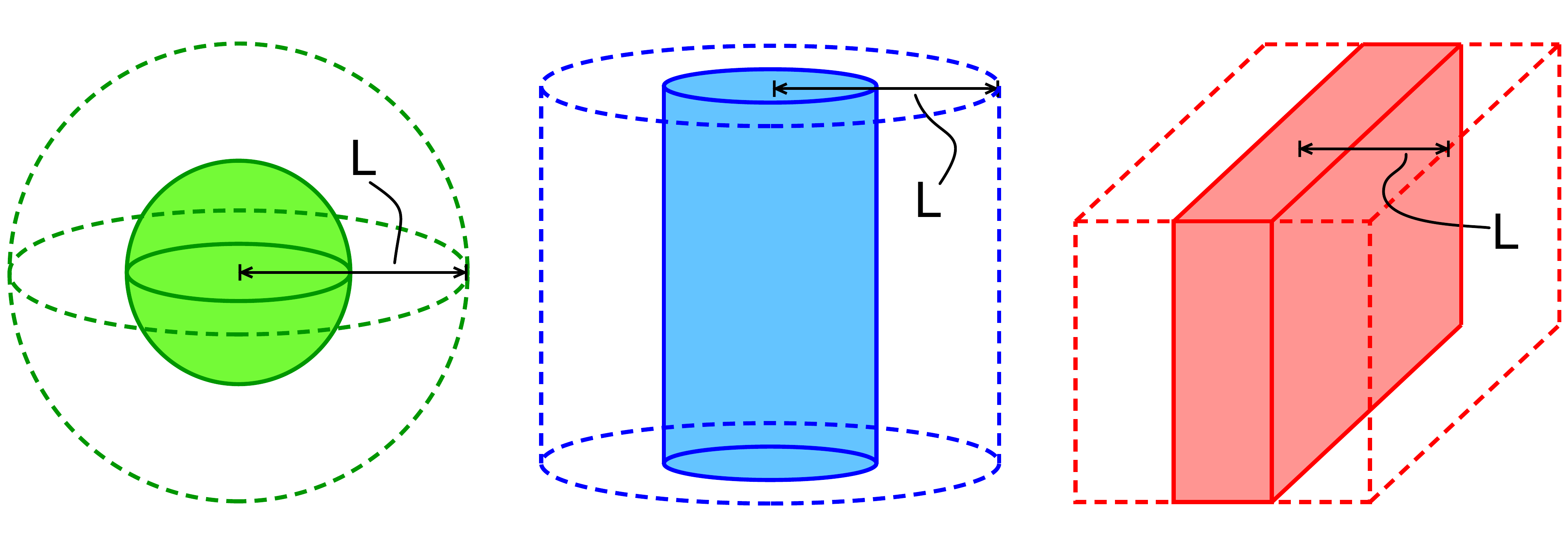}
\caption{Illustration of the geometries (bubbles, rods, slabs) of the 'pasta phases' considered in this paper. The inner regions (shaded) are occupied by one phase, and the rest of the unit cell (volume bounded by the dashed lines minus the shaded region) by the other phase. Together with the complementary structures for bubbles and rods, there are 5 different geometries in total. The profiles of the condensates and the electrostatic potential across the interfaces 
(here shown as sharp surfaces for illustrative purpose) are 
computed numerically and the free energy is minimized with respect to the 
size of the unit cell $L$. The unit cell is taken to have the same geometry as the
structure of the pasta, thus approximating the actual Wigner-Seitz cell. 
}
\label{fig:cells}
\vspace{-0.3cm}
\end{center}
\end{figure}

I will employ a nucleon-meson model  with a small 
explicit chiral symmetry breaking term \cite{Boguta:1982wr,Boguta:1986ha,Floerchinger:2012xd,Drews:2013hha}, which shows a first-order transition to an approximately chirally symmetric phase. This model was used in Ref.\ \cite{Fraga:2018cvr} to compute the surface tension for  isospin-symmetric nuclear matter at the chiral phase transition. I shall use the extension of the model to isospin-asymmetric matter \cite{Drews:2014spa} to construct mixed phases explicitly. This model has the obvious advantage of treating the phases on both sides of the transition on the same footing. In particular, location and properties of the chiral phase transition 
are determined by fixing all parameters to match known properties of nuclear matter at saturation, and interfaces between the two phases can be computed 
fully consistently, with the surface tension being a side result, not an input. The disadvantage is that the chirally restored phase is actually made of very light nucleons, not of quarks, and thus can only be a toy version of deconfined quark matter. Therefore, the quark-hadron pasta within this model is 
more accurately thought of as 'chiral pasta'. Obviously, it is desirable to have a unified description of quarks and hadrons which is more realistic on both sides of the transition. However, this is currently not possible from first principles and 
difficult to achieve even on the level of phenomenological models. (For 
a recent attempt in this direction using the gauge-gravity duality see Refs.\  \cite{Li:2015uea,Preis:2016fsp,Preis:2016gvf,BitaghsirFadafan:2018uzs}.)

I will work at zero temperature throughout the paper and employ the mean-field and no-sea approximations for the evaluation of the model. Furthermore, for the 
inhomogeneous phases I use the Thomas-Fermi approximation to integrate out the fermions and construct the mixed phase in the Wigner-Seitz approximation. 
In this approach, for a given geometry (bubbles, rods, slabs), the energetically preferred size of a unit cell is determined from the calculation of the interface profiles, see Fig.\ \ref{fig:cells} for an illustration. 
This profile calculation is done by solving simultaneously and numerically the Euler-Lagrange equations for the mesonic condensates together with the Poisson equation for the electrostatic potential.  A similar calculation was performed 
for the interface between vacuum and nuclear matter \cite{Maruyama:2005vb}, but, to the best of my knowledge, the present paper is the first such calculation in the context of the quark-hadron transition.

The paper is organized as follows. In Sec.\ \ref{sec:setup}, I introduce the model, explain the various approximations, derive the Euler-Lagrange equations, and comment on the model parameters and the specific choices I make for them. Sec.\ \ref{sec:mix1} contains an analysis of the homogeneous phases and serves to identify the regime in which mixed phases can potentially be found. The main results are presented Sec.\ \ref{sec:pasta}. After recapitulating the simple step-like approximation in Sec.\ \ref{sec:step} and explaining the numerical evaluation in Sec.\ \ref{sec:full}, I show and interpret the results for the chiral pasta phases in Sec.\ \ref{sec:results}. A summary and an outlook are given in Sec.\ \ref{sec:summary}. 
I use natural units $c=\hbar=k_B=1$ and Heaviside-Lorentz units for the gauge fields, where the 
elementary electric charge is $e=\sqrt{4\pi\alpha}\simeq 0.3$ with the fine structure constant $\alpha$. The convention for the metric tensor in Minkowski space is $g^{\mu\nu}=(1,-1,-1,-1)$.

\section{Setup}
\label{sec:setup}

\subsection{Nucleon-meson Lagrangian}

The starting point is the Lagrangian \cite{Drews:2014spa}
\bea \label{lag} \allowdisplaybreaks
{\cal L} &=& 
\bar{\psi}(i\gamma_\mu\partial^\mu + \gamma^0\hat{\mu})\psi + \frac{1}{2}\partial_\mu\sigma\partial^\mu\sigma + \frac{1}{4}\Tr[\partial_\mu\pi\partial^\mu\pi] \non[2ex] &&-\frac{1}{4}\omega_{\mu\nu}\omega^{\mu\nu}-\frac{1}{8}\Tr[\rho_{\mu\nu}\rho^{\mu\nu}]- \frac{1}{4}F_{\mu\nu}F^{\mu\nu} \non[2ex]
&&- {\cal U}(\sigma,\pi) +\frac{m_v^2}{2}\left(\omega_\mu\omega^\mu  +\frac{1}{2}\Tr[\rho_\mu\rho^\mu]\right) \non[2ex]
&&- \bar{\psi}\Big[g_\sigma(\sigma+ i\gamma^5\pi) + \gamma_\mu(g_\omega\omega^\mu+g_\rho\rho^\mu) \Big]\psi \, , 
\eea
where the traces are taken over isospin space. 
The nucleon field 
\be
\psi = \left(\begin{array}{c}\psi_n \\ \psi_p\end{array}\right) 
\ee
includes neutron and proton spinors with corresponding chemical potentials $\mu_n$, $\mu_p$, such that 
\be \label{muhat}
\hat{\mu}=\left(\begin{array}{cc} \mu_n & 0 \\ 0 & \mu_p \end{array}\right) = \left(\begin{array}{cc} \mu_B+\mu_I & 0 \\ 0 & \mu_B-\mu_I \end{array}\right)
\ee
is the chemical potential matrix in isospin space, with the baryon chemical potential $\mu_B$ and the isospin chemical potential $\mu_I$. The mesonic fields are $\sigma$, $\omega_\mu$, $\pi = \pi_a\tau^a$, and  $\rho_\mu = \rho_\mu^a\tau_a$, 
where $\tau_a$ ($a=1,2,3$) are the Pauli matrices, normalized such that $\Tr[\tau_a,\tau_b]=2\delta_{ab}$, $[\tau_a,\tau_b]=2i\epsilon_{abc}\tau_c$. 
The vector meson mass will be set to $m_v\simeq 782\, {\rm MeV}$. The kinetic terms of the vector mesons are defined through 
\begin{subequations}
\bea
\omega_{\mu\nu}&=& \partial_\mu\omega_\nu-\partial_\nu\omega_\mu \, , \\[2ex]
\rho_{\mu\nu}&=&  \partial_\mu\rho_\nu-\partial_\nu\rho_\mu+i\frac{g_\rho}{2}[\rho_\mu,\rho_\nu] \, ,
\eea
\end{subequations}
and $F_{\mu\nu} = \partial_\mu A_\nu-\partial_\nu A_\mu$ is the electromagnetic field strength tensor with the electromagnetic gauge field $A_\mu$.  The potential for the sigma and pion fields is taken to be 
\be \label{U}
{\cal U}(\sigma,\pi) = \sum_{n=1}^4 \frac{a_n}{n!} \frac{(\chi^2-f_\pi^2)^n}{2^n}-\epsilon(\sigma-f_\pi) \, , 
\ee
with  
\be
\chi^2\equiv \sigma^2+\frac{1}{2}\Tr[\pi^2] = \sigma^2 + \pi_1^2+\pi^2_2+\pi_3^2 \, , 
\ee
and with the pion decay constant $f_\pi\simeq 93\, {\rm MeV}$ and the model parameters $a_1,a_2,a_3,a_4$ and
$\epsilon$. By requiring that in the vacuum $\langle\chi\rangle=f_\pi$, the pion mass term fixes 
$a_1 = m_\pi^2$ with the pion mass $m_\pi\simeq 139\, {\rm MeV}$, as well as the explicit chiral symmetry breaking term $\epsilon = m_\pi^2 f_\pi$. In addition to these parameters, the model 
also contains the coupling constants $g_\sigma$, $g_\omega$, $g_\rho$, which 
determine the nucleon-meson coupling via Yukawa interactions. The choice for the 6 remaining free parameters will be discussed in Sec.\ \ref{sec:parameters}.  

\subsection{Free energy density}

To facilitate the numerical evaluation of the pasta phases, I make use of several approximations. Employing the mean-field approximation, I neglect all meson fluctuations, such that the meson contributions arise only through the mesonic condensates $\bar{\sigma}$, $\bar{\omega}$, and $\bar{\rho}$. Here, $\bar{\sigma}\equiv \langle\sigma\rangle$ plays the role of the chiral condensate, giving rise to the nucleon mass. The omega condensate is the expectation value of the temporal component of the omega vector field, $\bar{\omega}\equiv \langle\omega_0\rangle$, and the rho condensate corresponds to the temporal component of the third isospin component of the rho vector field, $\bar{\rho}\equiv \langle\rho_0^3\rangle$. The condensates $\bar{\omega}$ and $\bar{\rho}$ act as effective contributions to the baryon and isospin chemical potentials. The resulting mean-field Lagrangian is equivalent to 
a Lagrangian of non-interacting fermions, with all effects of the interactions absorbed in the effective mass and effective chemical potentials. The ansatz for the condensates does not include the possibility of a chiral density wave, where the 
chiral condensate oscillates spatially between the $\sigma$ and $\pi_3$ directions
with fixed modulus $\chi$ \cite{Heinz:2013hza}. The chiral density wave or related, more complicated, 
inhomogeneous structures are expected to appear in the vicinity of a
first-order chiral phase transition \cite{Buballa:2014tba}, 
even without relaxing the 
local neutrality constraint to a global one, which is necessary for the mixed phases discussed here. I leave the study of the interplay between the chiral density wave and chiral pasta for the future. For simplicity, I will also drop the vacuum contribution, i.e., I work 
in the so-called no-sea approximation. Since the model is renormalizable, it is straightforward to include this contribution, which would introduce the renormalization scale as an additional parameter. However, since the model is 
of phenomenological nature to begin with, there is no guarantee that the vacuum terms would make the results more realistic, although there are examples where it does so \cite{Heide:1991wn}. There are cases where the 
sea contribution makes an obvious {\it qualitative} difference, for instance in 
the presence of a magnetic field \cite{Haber:2014ula}, but such an effect is not expected here. 

In order to describe inhomogeneous mixed phases, the condensates $\bar{\sigma}$, $\bar{\omega}$, and $\bar{\rho}$ must be allowed to vary in space. (There is no time dependence since I am only interested in equilibrium configurations.) Within the Thomas-Fermi approximation, this spatial dependence is neglected when integrating over the fermionic fields to compute the partition function and thus the 
free energy density $\Omega$ of the system. As a consequence, the resulting free energy density has the same form as in the homogeneous case with all mesonic condensates allowed to be inhomogeneous and with the addition of the mesonic gradient terms. This approximation is expected to be valid for slowly varying condensates. 

Within these approximations, the calculation of the 
free energy density follows the standard procedure, and at zero temperature one obtains 
\bea \label{Omega}
\Omega 
&=&-\frac{(\nabla\bar{\omega})^2}{2}-\frac{(\nabla\bar{\rho})^2}{2}+\frac{(\nabla\bar{\sigma})^2}{2}+\frac{(\nabla\mu_e)^2}{2e^2}\non[2ex]
&&+\,\Omega_N(\bar{\sigma},\bar{\omega},\bar{\rho}) +\Omega_{\ell} \, . 
\eea
Here, the Coulomb energy density is given by the electric field $\vec{E} = -\nabla\phi$, where the scalar electric potential $\phi$ is related to the 
electron chemical potential $\mu_e$ by $\phi=\mu_e/e$. In addition to the terms
coming from the nucleon-meson Lagrangian (\ref{lag}), I have added a leptonic contribution from non-interacting, approximately massless electrons with chemical potential $\mu_e$ and non-interacting muons with chemical potential $\mu_\mu$ and mass $m_\mu \simeq 106\, {\rm MeV}$,
\be \label{Omell}
\Omega_\ell=-\frac{\mu_e^4}{12\pi^2}-p(\mu_\mu,m_\mu) \,, 
\ee
where I have introduced the zero-temperature pressure of a non-interacting spin-$\frac{1}{2}$ fermion species with chemical potential $\mu$ and mass $m$, 
\bea
p(\mu,m)&\equiv& \frac{\Theta(\mu-m)}{8\pi^2} \non[2ex]
&&\hspace{-1cm}\times \left[\left(\frac{2}{3}k_F^3-m^2k_F\right)\mu+m^4\ln\frac{k_F+\mu}{m}\right] \, , 
\eea
with the Fermi momentum
\be
k_F = \sqrt{\mu^2-m^2} \, .
\ee
For $m=0$ the pressure reduces to the simple form used for the 
electrons in Eq.\ (\ref{Omell}). The nucleonic contribution in Eq.\ (\ref{Omega}) is  
\bea\label{Omega0}
\Omega_N(\bar{\sigma},\bar{\omega},\bar{\rho}) 
&\equiv& -\frac{1}{2}m_v^2(\bar{\omega}^2+\bar{\rho}^2)+{\cal U}(\bar{\sigma})\non[2ex] &&-p(\mu_n^*,M)-p(\mu_p^*,M) \, , 
\eea
where ${\cal U}(\bar{\sigma})$ is the potential (\ref{U}) with all fields replaced by their expectation values, i.e., $\chi=\bar{\sigma}$, where 
\begin{subequations} \label{mustar}
\bea
\mu_n^* &\equiv & \mu_n-g_\omega\bar{\omega}-g_\rho\bar{\rho} \, , \\[2ex]
\mu_p^*&\equiv&  \mu_p-g_\omega\bar{\omega}+g_\rho\bar{\rho} 
\eea
\end{subequations}
are the effective nucleon chemical potentials,
and where
\be
M \equiv g_\sigma \bar{\sigma}  
\ee
is the effective nucleon mass. 

For low temperatures, where the neutrino mean free path is much larger than the size of the neutron star, equilibrium with respect to the weak processes 
$p+\ell\to n+\nu_\ell$, $n\to p+\ell+\bar{\nu}_\ell$ where $\ell= e,\mu$ is either an electron or a muon, and $e\to \mu +\bar{\nu}_\mu+\nu_e$, $\mu\to e+\bar{\nu}_e+\nu_\mu$ 
yields the following relations between the nucleonic and leptonic chemical potentials\footnote{Notice that these relations hold for the actual chemical potentials $\mu_n$ and $\mu_p$, not for the auxiliary quantities $\mu_n^*$ and $\mu_p^*$. The reason is that the nucleonic single-particle energies at the Fermi momentum are $\mu_n$ and $\mu_p$, not $\mu_n^*$ and $\mu_p^*$.}, 
\be \label{weakeq}
\mu_p+\mu_e=\mu_n \, , \qquad \mu_\mu = \mu_e \, . 
\ee
I will use these constraints to eliminate $\mu_p$ and $\mu_\mu$, such that the only remaining chemical potentials are $\mu_n$ and $\mu_e$.

\subsection{Euler-Lagrange equations}

From the free energy density (\ref{Omega}) one can now straightforwardly derive the Euler-Lagrange equations, 
\begin{subequations}\label{dOm}
\bea
\nabla^2 \bar{\sigma} &=& \frac{\partial \Omega_N}{\partial\bar{\sigma}}
= \frac{\partial {\cal U}}{\partial\bar{\sigma}} + g_\sigma n_s \, ,  \label{eom1} \\[2ex] 
\nabla^2 \bar{\omega} &=& 
-\frac{\partial\Omega_N}{\partial\bar{\omega}} = m_v^2\bar{\omega} - g_\omega n_B\, , \label{eqomega} \\[2ex]
\nabla^2 \bar{\rho} &=& 
-\frac{\partial\Omega_N}{\partial\bar{\rho}} = m_v^2\bar{\rho} - g_\rho n_I \, , \label{eqrho} \\[2ex]
\frac{\nabla^2 \mu_e}{e^2} &=& 
-\left(\frac{\partial\Omega_N}{\partial\mu_e}+\frac{\partial\Omega_\ell}{\partial\mu_e}\right) = -q  \, . \label{neutrality}
\eea
\end{subequations} 
The last equation is the Poisson equation, which can also be written as 
\be
\nabla^2\phi=-\rho=-eq\, , 
\ee
where $\rho$ is the electric charge density, and 
\be \label{qn}
q\equiv n_p-n_e-n_\mu \, , 
\ee
such that positive $\rho$ and $q$ correspond to an excess of proton over lepton charges. 
The scalar density is
\bea
&&n_s =  \Theta(\mu_{n}^*-M) \frac{M}{2\pi^2}\left(k_{F,n}\mu_{n}^*-M^2\ln\frac{k_{F,n}+\mu_{n}^*}{M}\right)
\non[2ex]
&&+\Theta(\mu_{p}^*-M) \frac{M}{2\pi^2}\left(k_{F,p}\mu_{p}^*-M^2\ln\frac{k_{F,p}+\mu_{p}^*}{M}\right) \, , 
\eea
where $k_{F,n}$ and $k_{F,p}$ are the neutron and proton Fermi momenta. 
The baryon and isospin densities are $n_B=n_n+n_p$, $n_I=n_n-n_p$ with 
the neutron and proton densities 
\bea \label{nnnp}
 n_n&=& \Theta(\mu_{n}^*-M)\frac{k_{F,n}^3}{3\pi^2} \, , \quad n_p = \Theta(\mu_{p}^*-M)\frac{k_{F,p}^3}{3\pi^2} \, , \hspace{0.3cm}
\eea
and the electron and muon densities are
\bea
 n_e&=& \frac{\mu_e^3}{3\pi^2} \, , \quad n_\mu = \Theta(\mu_{\mu}-m_\mu)\frac{(\mu_\mu^2-m_\mu^2)^{3/2}}{3\pi^2} \, .
\eea
The four coupled second-order differential equations (\ref{dOm}) will later be solved numerically 
for the four functions $\bar{\sigma}(\vec{r})$, $\bar{\omega}(\vec{r})$, $\bar{\rho}(\vec{r})$, $\bar{\mu}_e(\vec{r})$.  
In the homogeneous limit, Eqs.\ (\ref{dOm}) become coupled algebraic equations. Then, Eqs.\ (\ref{eom1}) -- (\ref{eqrho}) are the stationarity equations of the free energy with respect to the variables $\bar{\sigma}$, $\bar{\omega}$, $\bar{\rho}$, and Eq.\ (\ref{neutrality}) accounts for electric charge neutrality. 

\subsection{Parameter values}
\label{sec:parameters}

It remains to fix the parameters $g_\sigma$, $g_\omega$, $g_\rho$, $a_2$, $a_3$, $a_4$. The matching procedure is almost the same as in Ref.\ \cite{Fraga:2018cvr}, where only isospin-symmetric matter was discussed. We first set $g_\sigma = 10.097$, which follows from the nucleon mass in the vacuum $m_N = 939\, {\rm MeV}$ and the relation $m_N = g_\sigma f_\pi$. The remaining 5 parameters depend, in a coupled way, on saturation properties of symmetric nuclear matter which are not all known to high accuracy. I use the saturation density $n_0=0.153\, {\rm fm}^{-3}$ and the binding energy $E_B=-16.3\, {\rm MeV}$, both known to good accuracy. Furthermore, I need the incompressibility $K \simeq (200-300)\, {\rm MeV}$ \cite{Blaizot:1995zz,Youngblood:2004fe}, the effective Dirac mass at saturation $M_0\simeq (0.7-0.8)\, m_N$ \cite{1989NuPhA.493..521G,Johnson:1987zza,Li:1992zza,glendenningbook,Jaminon:1989wj,Furnstahl:1997tk}, and the asymmetry energy $S\simeq (30.2-33.7)\, {\rm MeV}$ \cite{Danielewicz:2013upa,Lattimer:2014sga}. For definiteness, I set 
$S=32\, {\rm MeV}$, as in Ref.\ \cite{Drews:2014spa}, and $M_0=0.75\, m_N$. One can easily express the coupling constant $g_\omega$ in an analytical form in terms of the experimental values, 
\be
g_\omega^2 = \frac{m_v^2}{n_0}\left(\mu_0-\sqrt{k_{F,0}^2+M_0^2}\right) \simeq 89.6179 \, ,
\ee
where $\mu_0=m_N+E_B=922.7\, {\rm MeV}$ is the baryon chemical potential at the onset of symmetric nuclear matter, and where 
\be
k_{F,0} = \left(\frac{3\pi^2 n_0}{2}\right)^{1/3} 
\ee
is the Fermi momentum of symmetric nuclear matter at saturation. An analogous expression can be derived for the coupling constant $g_\rho$, 
\be \label{grho}
g_\rho^2 = \frac{3m_v^2\pi^2}{k_{F,0}^3}\left(S-\frac{k_{F,0}^2}{6\sqrt{k_{F,0}^2+M_0^2}}\right) \simeq 17.7737 \, . 
\ee
This relation is derived in appendix \ref{app:sym}. 
The remaining parameters $a_2$, $a_3$, $a_4$ are then determined from the stationarity equation with respect to $\bar{\sigma}$, the condition that nuclear matter at the onset have zero pressure, and that the incompressibility $K$ at saturation have a certain value, where the expression for $K$ is also derived in  appendix \ref{app:sym}, see Eq.\ (\ref{K}) for the final result. These three conditions are coupled and have to be solved numerically. Due to the experimental uncertainty in $K$, and in order to look for qualitatively different scenarios, I shall first treat $K$ as free parameter. For the actual numerical calculation of the mixed phases I will then focus on the single parameter value $K=252\, {\rm MeV}$, for which, with the given values for $n_0$, $E_B$, $M_0$, $S$, one finds $a_2=45.4695$, $a_3=-8.42367\times 10^{-3}\, {\rm MeV}^{-2}$, $a_4=4.97732\times 10^{-5}\, {\rm MeV}^{-4}$.

\section{Identifying mixed phase regions}
\label{sec:mix1}

Before calculating the profiles of the condensates in various mixed phase geometries, one needs to identify the regions in the phase diagram where the 
mixed phases are candidates to replace the homogeneous solutions. At zero temperature, the only independent thermodynamic variable is the neutron chemical potential $\mu_n$, i.e., for a given set of model parameters, this amounts to identifying a range in $\mu_n$ around the first-order phase transition. 

The model has three different phases: the baryonic vacuum (V), chirally broken nuclear matter (N), and the chirally restored phase, which I loosely interpret as quark matter (Q), as discussed in the introduction. Since chiral symmetry is explicitly broken, N and Q are not distinguished by symmetry and in principle there can be a smooth transition between them. However, in the parameter range of interest it turns out that they are separated by a first-order transition, and hence it does make sense to label them as distinct phases.
For isospin-symmetric nuclear matter, there is a first-order transition from V to N, the 'baryon onset', to whose properties the parameters of the model are fitted, as  discussed above. Then, at larger chemical potentials, there is a transition to Q. This transition is of first order unless the incompressibility assumes unrealistically large values, in which case this transition becomes a smooth crossover \cite{Fraga:2018cvr}. 

\begin{figure} [t]
\begin{center}
\includegraphics[width=\columnwidth]{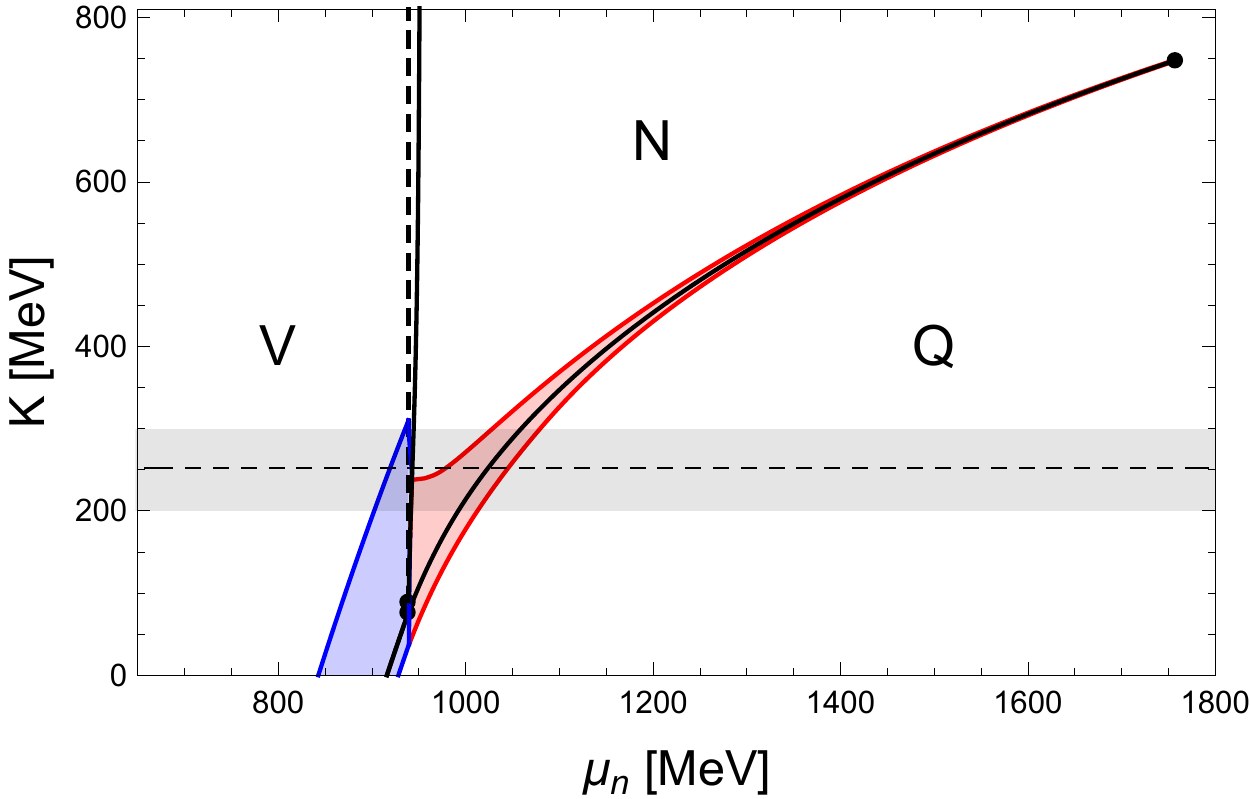}
\caption{Mixed phase regions in the plane of incompressibility at saturation $K$
and neutron chemical potential $\mu_n$, without Coulomb effects and surface tension for the nuclear-quark (NQ, red shaded region) and vacuum-quark (VQ, blue shaded region) mixed phases. Thick black curves are first-order (solid) and second-order (dashed) phase transitions between homogeneous phases. 
The baryon onset is second order almost everywhere, and there is a first-order transition within the nuclear matter phase just after the onset. The grey band indicates the experimental uncertainty for $K$, and the horizontal thin dashed line shows the value of $K$ used in all following results. 
}
\label{fig:mixed1}
\end{center}
\end{figure}

\begin{figure} [t]
\begin{center}
\includegraphics[width=\columnwidth]{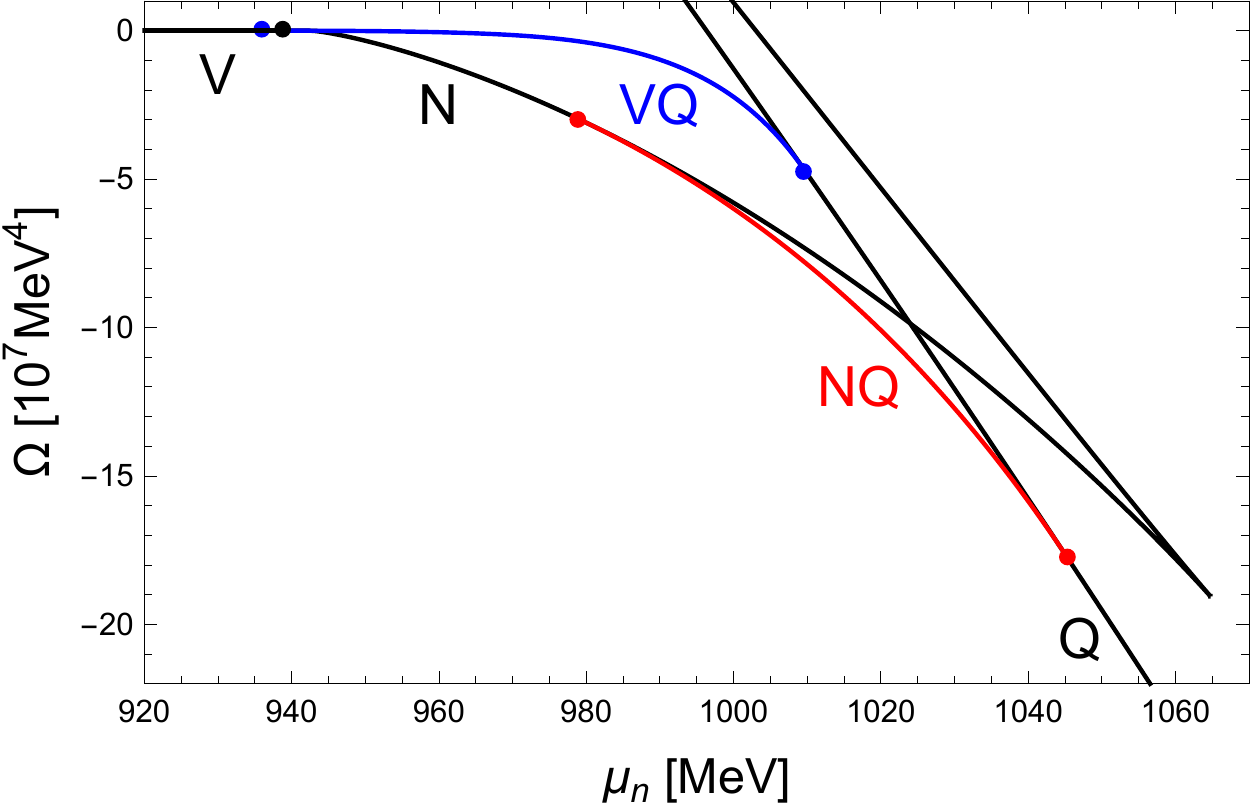}
\caption{Free energy density as a (multi-valued) function of the neutron chemical potential  for locally neutral phases N, V, Q (black) and mixed, globally neutral, phases VQ (blue) and  NQ (red),  neglecting Coulomb and surface effects, and using $K=252\, {\rm MeV}$. The main point of the paper is to include Coulomb and surface effects and thus determine the fate of the NQ curve. 
}
\label{fig:omega}
\end{center}
\end{figure}

\begin{figure} [t]
\begin{center}
\includegraphics[width=\columnwidth]{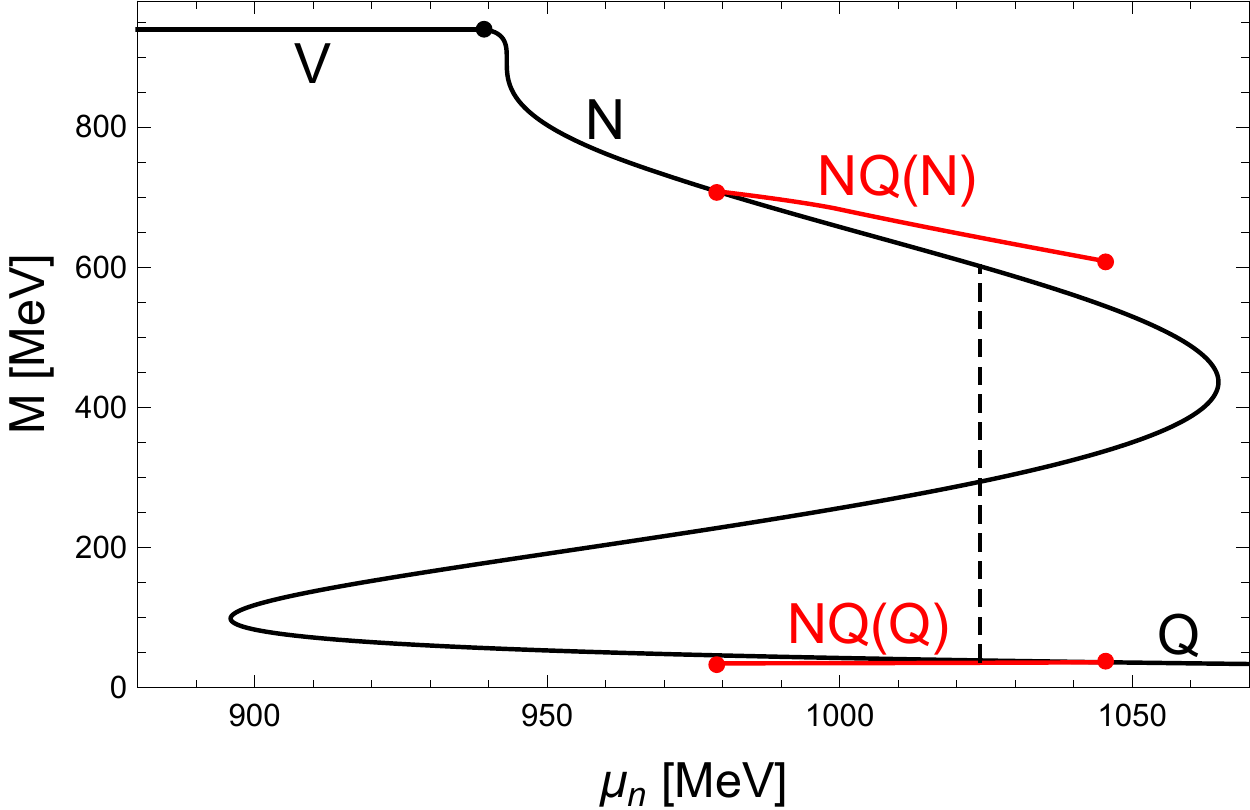}

\includegraphics[width=\columnwidth]{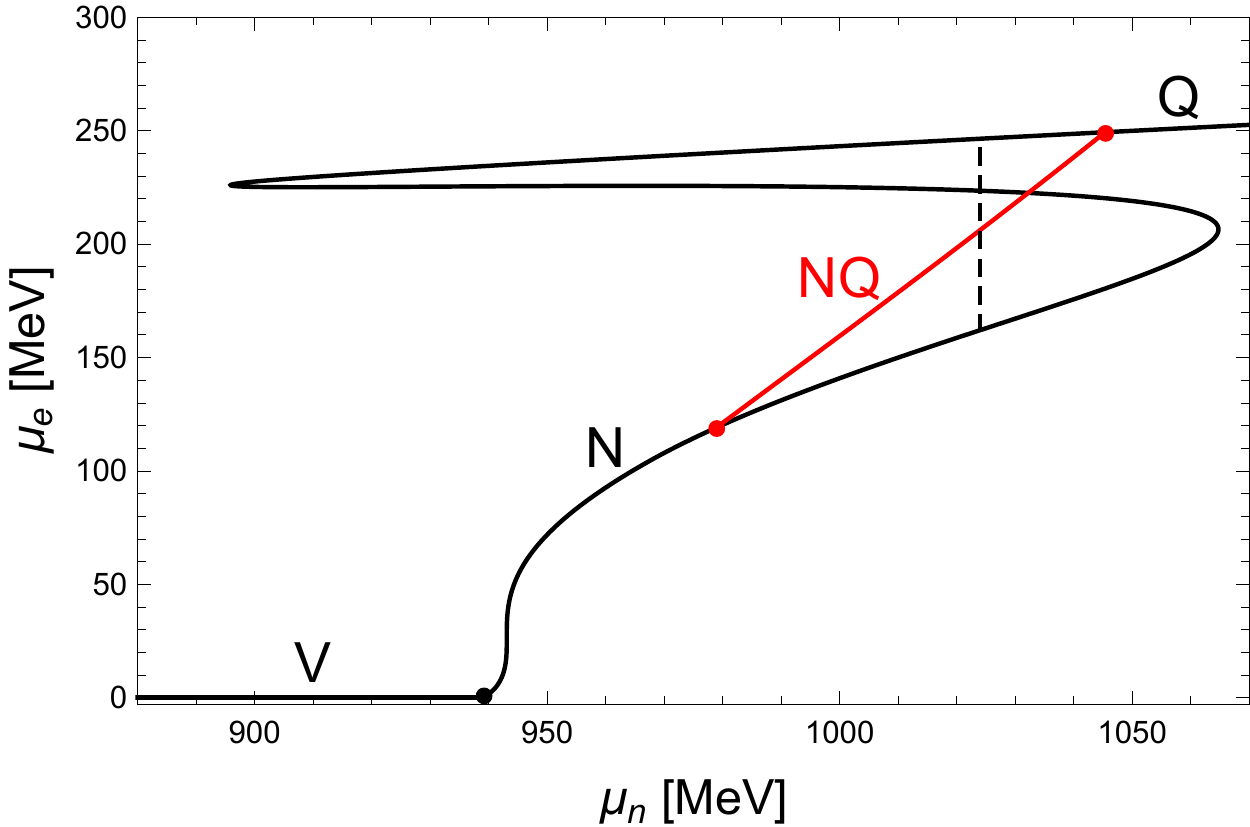}
\caption{Effective nucleon mass $M$ and electron chemical potential $\mu_e$ for locally neutral phases (black curves, the NQ first-order phase transition is indicated by the vertical dashed lines) and the NQ mixed phases without Coulomb and surface effects (red curves), for the same parameters as in Fig.\ \ref{fig:omega}. The jump from large to almost vanishing mass $M$ is characteristic for the chiral phase transition. While N and Q phases have different values of $M$ in the mixed phase, there is only a single value of $\mu_e$. 
}
\label{fig:MMue}
\end{center}
\end{figure}

For isospin-asymmetric, electrically neutral, beta-equilibrated  matter the overall phase structure is similar, but there are some differences, as Fig.\ \ref{fig:mixed1} shows. The black curves in this figure correspond to the phase transitions between the homogeneous phases. There is a small region of unrealistically small $K$ where a direct transition from V to Q occurs. In this region, nuclear matter is only metastable and never the state of lowest free energy for any neutron chemical potential. (For symmetric nuclear matter, this regime is larger, extending into the region of realistic values of $K$ \cite{Fraga:2018cvr}.) Then, there is a very small regime, hardly visible in the plot, where the onset to neutral baryonic matter is first order. For most of the parameter range -- including all realistic values of $K$ -- the baryon onset is second order, shown by the dashed line. The solid line just next, and almost parallel, to the dashed line indicates a (weak) first-order phase transition within the nuclear matter phase.  I am not aware of any other model showing such a first-order transition, and possibly it is an artifact of the model. In any case, this transition will not be relevant for the main results of this paper. A more detailed view of the homogeneous phases, for a particular value of $K$, can be found in Figs.\ \ref{fig:omega} and \ref{fig:MMue}, where the free energy density, the effective nucleon mass, and the electron chemical potential are plotted. All these results are obtained by solving the Euler-Lagrange equations (\ref{eom1}) -- (\ref{eqrho}) without gradient terms together with the {\it local}  neutrality constraint from Eq.\ (\ref{neutrality}). Figures \ref{fig:mixed1} -- \ref{fig:MMue} also contain results for the VQ and NQ mixed phases in the absence of surface and Coulomb effects, where Eq.\ (\ref{neutrality}) is turned into a {\it global} neutrality condition, as I explain now.

Mixed phases can appear in the vicinities of the first-order phase transitions if one allows for spatial regions that are electrically charged, with the overall charge of the system being zero.  In the simplest approximation, Coulomb energy and surface tension are neglected, and the mixed phase is characterized solely by the volume fraction of one of the phases $\chi\in [0,1]$. In particular, in this approximation, the specific spatial structure of the system is completely irrelevant. Adding Coulomb and surface energies increases the free energy of the mixed phase, and thus this simple  approximation overestimates the stability of the mixed phases and gives an upper limit for the range in $\mu_n$ in which the mixed 
phase is favored. The calculation is done as follows. Suppose phases 1 and 2 occupy volume fractions $\chi$ and $1-\chi$, respectively. 
Then, Eqs.\ (\ref{eom1}) -- (\ref{eqrho}) for each phase (6 equations in total) together with  the condition of global charge neutrality,  
\be
0 = \chi q_1+(1-\chi) q_2 \, , 
\ee
with the charge densities for the two phases defined via Eq.\ (\ref{qn}), and the equality of the free energies densities,
\be
\Omega_1 = \Omega_2 \, , 
\ee
are solved simultaneously for the 8 scalar variables $\bar{\sigma}_1, \bar{\omega}_1, \bar{\rho}_1, \bar{\sigma}_2, \bar{\omega}_2, \bar{\rho}_2, \mu_e, \chi$ for a given neutron chemical potential $\mu_n$. As a result, one finds a range in $\mu_n$  where $\chi$ decreases continuously from $\chi=1$ at the lower boundary to $\chi=0$ at the upper boundary.
In other words, as $\chi$ goes from 1 to 0, one moves from a pure system in phase 1 through a mixture of both phases up to the pure phase 2. The resulting free energy densities for the VQ and NQ mixed phases are shown in Fig.\ \ref{fig:omega}. The VQ phase is shown for all $\mu_n$ where it exists, even though in a large part of this regime it is not the state of lowest free energy. 
Here and throughout the paper the VN mixed phase is ignored for simplicity.
The situation becomes even more complicated if the mixed phase in the vicinity of the first-order transition within the N phase is constructed, made of nuclear matter with two different densities. This mixed phase is not included in the analysis either. In any case, the focus of this paper is on the NQ mixed phase, which is stable throughout the regime where it is a solution. 
The corresponding curves for the effective masses and the electron chemical potential in Fig.\ \ref{fig:MMue} are therefore shown only for the NQ phase.

To get an idea of the potential importance of mixed phases in this model, I have determined the range for the mixed phases for all values of the incompressibility $K$. 
Of course there are experimental constraints for $K$ and one might argue that large parts of the parameter space are therefore not interesting. However, one needs to keep in mind that the model used here is of phenomenological nature. Therefore, it is useful to scan the parameter space and look for interesting qualitative features regarding mixed phases. While certain features might not be realized in the present model for realistic values of $K$, these features might be realized in other models, or in improved versions of this model, or in QCD. In this spirit, one might also add variations in $M_0$ and $S$ to extend the result presented here, as it was done, varying $K$ {\it and} $M_0$, for isospin-symmetric nuclear matter in Ref.\ \cite{Fraga:2018cvr}. The result for 
fixed $M_0$ and $S$, but $K$ unconstrained, is shown in the form of the red and blue shaded areas in Fig.\ \ref{fig:mixed1}. In contrast to Fig.\ \ref{fig:omega}, the VQ mixed phase is now indicated only in the regime where it has lower free energy than all homogeneous phases and the NQ mixed phase. One observes that the NQ mixed phase region in the vicinity of the chiral phase transition becomes smaller as $K$ is increased, eventually vanishing for very large $K$, where the transition becomes a smooth crossover. 
Interestingly, the results show the possibility of a direct transition from the VQ mixed phase to the NQ mixed phase. Such a behavior is suggested for $K \simeq (30-240)\, {\rm MeV}$. If realized in QCD, this might for instance be relevant for quark stars, which are usually thought of as having an outer layer  
where strangelets are immersed in an electron gas, i.e., they have a crystalline VQ crust and a homogeneous Q body. Figure \ref{fig:mixed1} suggests that stars of the form VQ-NQ-Q are conceivable. In other words, nuclear matter might find its way into a quark star through a mixed phase that  exists in a certain layer of the star. Moreover, the transition from the VQ to the NQ mixed phase raises the question whether a VQN mixed phase exists in the transition region, i.e., a phase in which baryonic vacuum, nuclear matter, and quark matter  coexist in spatially separated regions. These intriguing questions are beyond the scope of the present paper but deserve further studies,  be it in the present model or in another approach.  

Here I  proceed with the parameter set of Figs.\ \ref{fig:omega} and \ref{fig:MMue}, which shows the 'standard' form of the NQ mixed phase -- surrounded by pure nuclear matter N at lower density and by pure quark matter Q at higher density -- and will no longer be concerned with the VQ mixed phase and any complications that are connected with it.

\section{Pasta phases}
\label{sec:pasta}

Whether chiral pasta phases are preferred or not is determined by the 
total free energy per unit volume, which is given by the spatial integral over the free energy density (\ref{Omega})
\bea \label{F}
F &=&  w_C+\frac{1}{V}\int d^3 \vec{r}\left[\frac{m_v^2}{2}(\bar{\rho}^2+\bar{\omega}^2)-\frac{\bar{\sigma}}{2}\frac{\partial {\cal U}}{\partial \bar{\sigma}} \right.\non[2ex]
&&\left.-\frac{1}{2}(g_\omega\bar{\omega}\, n_B+g_\rho\bar{\rho}\,  n_I+g_\sigma\bar{\sigma} \, n_s)+\Omega_N+\Omega_\ell \right] \, , \hspace{0.5cm}
\eea
where partial integration and the Euler-Lagrange equations (\ref{eom1}) -- (\ref{eqrho}) have been used, where the surface terms are dropped due to $\nabla\bar{\omega}=\nabla\bar{\rho}=\nabla\bar{\sigma}=0$ at the boundaries, and 
where the Coulomb energy per unit volume has been separated,
\be \label{WC}
w_C = \frac{1}{2V}\int d^3\vec{r}\, E^2 \, , 
\ee
with the electric field $\vec{E} = \nabla\mu_e/e$.  One can also compute $F$ by directly integrating over $\Omega$ from Eq.\ (\ref{Omega}), 
without rewriting the 
gradient terms with the help of partial integration. It is a good numerical check, however,  to use both expressions because 
Eq.\ (\ref{F}) makes explicit use of the Euler-Lagrange equations. This check can
thus be used to confirm that these equations are solved to good accuracy. 

For the determination of the preferred pasta structure, $F$ is the only relevant quantity and there is no need to extract the surface tension $\Sigma$. Nevertheless, it is useful to compute $\Sigma$ because, firstly, it can be used  as an input for simple approximations that do not rely on the knowledge of the exact spatial profiles and, secondly, it is interesting to compare the result to the surface tension of isospin-symmetric matter. 
The surface tension is defined 
by the difference in free energies between a domain wall configuration and the corresponding homogeneous 
configuration, 
\bea \label{Sigma}
\Sigma &=& \int_{x_{01}}^{x_{02}} dx\Bigg[-\frac{1}{2}\left(\frac{d\bar{\omega}}{dx}\right)^2-\frac{1}{2}\left(\frac{d\bar{\rho}}{dx}\right)^2+\frac{1}{2}\left(\frac{d\bar{\sigma}}{dx}\right)^2\non[2ex]
&&+\,\Omega_N+\Omega_\ell - (\Omega_N+\Omega_\ell)_0\Bigg]  \, ,
\eea
where $x_{01}$ and $x_{02}$ are the boundaries of the one-dimensional structure, and $(\Omega_N+\Omega_\ell)_0\equiv (\Omega_N+\Omega_\ell)_{x=x_{01}}=(\Omega_N+\Omega_\ell)_{x=x_{02}}$ is the free energy of either phase at the boundary. Note that the definition (\ref{Sigma}) 
does not include the Coulomb energy $w_C$. 

I will work within the Wigner-Seitz approximation, which reduces the problem to a single unit cell instead of the full crystalline structure. Moreover, as already shown in Fig.\ \ref{fig:cells}, the 
unit cells are assumed to have the form of the pasta structure itself, i.e., spherical for bubbles, cylindrical for rods, and rectangular for slabs, rather than using the actual Wigner-Seitz cells, which are the unit cells of the specific lattice structure (for instance, cubic, face-centered cubic). As a consequence, this approximation does not distinguish between different lattice structures of the bubbles or rods. The advantage is of course that the calculation becomes effectively one-dimensional for all three geometries.

\subsection{Step-like approximation}
\label{sec:step}

The Wigner-Seitz approximation (with conveniently shaped unit cells) can be further simplified if, additionally, the profiles of the condensates are assumed to be step-like. This approximation is well known and frequently used in the literature. Nevertheless, I will briefly recapitulate it because it serves as an introduction to the basic concepts of the chiral pasta phases, which is useful for the more 
complete calculation explained subsequently. Also, I will use the results of the step-like approximation later as a comparison to the full numerical result. 

In the step-like approximation, the sketches in Fig.\ \ref{fig:cells} can be taken literally: the interfaces that separate the two phases from each other are assumed to be sharp surfaces. The condensates and the electron chemical potential assume the values calculated in the previous section, as shown in Fig.\ \ref{fig:MMue}. Therefore, for each $\mu_n$ for which a mixed phase is conceivable according to Figs.\ \ref{fig:omega} and \ref{fig:MMue}, there are two electric charge densities, one for each constituent phase, which I will denote by $\rho_1$ and $\rho_2$, and a volume fraction $\chi$. This is the input from the microscopic calculation. One can now choose a certain geometry, and a simple exercise in electrostatics 
gives the Coulomb energy per unit volume. By assumption, the electrostatic potential, which is generated by the step-like charge distribution and which adds to the electron chemical potential, does not backreact on the values of the condensates. The calculation of the Coulomb energy is presented in 
appendix \ref{app:coulomb}, which leads to the well-known result \cite{glendenningbook}
\be \label{WCV}
w_C = \frac{(\rho_1-\rho_2)^2}{2}L_0^2\chi f_d(\chi) \, ,
\ee
where $L_0$ is the width of the inner region of the Wigner-Seitz cell (shaded regions in Fig.\ \ref{fig:cells}), i.e., the radius of the bubble or the radius of the rod or (half) the width of the slab, which is related to the volume fraction by 
\bea \label{chi}
\chi = \left(\frac{L_0}{L}\right)^d \, , 
\eea
where $L$ is the width of the Wigner-Seitz cell (see also Fig.\ \ref{fig:cells}), and $d$ is the co-dimension of the structure, i.e., $d=3$ for bubbles, $d=2$ for rods, and $d=1$ for slabs. Moreover,  
\be \label{fd}
f_d(\chi) = \left\{\begin{array}{cc} \displaystyle{\frac{(\chi-1)^2}{3\chi}} & \mbox{for} \;\; d=1 \\[2ex]
\displaystyle{\frac{\chi-1-\ln\chi}{4}} & \mbox{for} \;\; d=2 \\[2ex]\displaystyle{\frac{2+\chi-3\chi^{1/3}}{5}} & \mbox{for} \;\; d=3 
\end{array}\right. \, .
\ee
The Coulomb energy is obviously an energy  cost for creating spatial regions that are electrically charged. As Eq.\ (\ref{WCV}) confirms, this cost increases as the Wigner-Seitz cell becomes larger at a fixed volume fraction $\chi$. (Increasing the width of the Wigner-Seitz cell at fixed $\chi$ is equivalent to increasing $L_0$ at fixed $\chi$.) A counteracting  effect comes from the surface energy
 per unit volume,
\be \label{WSV}
w_S=\frac{d\Sigma \chi}{L_0} \, , 
\ee
which {\it decreases} as the size of the Wigner-Seitz  cell is increased because a larger 
Wigner-Seitz cell means fewer surfaces per volume where the condensates have to interpolate between the different phases, which costs energy. 
For a given $\chi$, the sum of Coulomb and surface energies 
\be \label{dF0}
\Delta F = w_C+w_S
\ee
 always has a minimum value at a certain nonzero and finite  $L_0$ because the Coulomb energy goes like $L_0^2$ and the surface energy goes like $1/L_0$. This will be different in the full calculation, for instance due to screening effects. Minimizing $\Delta F$ 
with respect to $L_0$ at fixed $\chi$ and inserting the result back into $\Delta F$ gives
\be \label{DeltaF}
\Delta F = \frac{3}{2}(\rho_1-\rho_2)^{2/3}\Sigma^{2/3}\chi[d^2f_d(\chi)]^{1/3} \, .
\ee
In order to determine the critical values of $\chi$ at which the geometry of the mixed phase changes, one simply has to evaluate the function $d^2f_d(\chi)$ 
for the five possible phases that arise from $d=1,2,3$ and from replacing $\chi\to 1-\chi$ in the cases $d=2,3$. The latter accounts for the fact that the bubbles or rods are made of phase 1 immersed in phase 2 or vice versa. For slabs the situation is symmetric and no additional information is obtained by the replacement $\chi\to 1-\chi$. One finds that the geometry changes from bubbles to rods to slabs to complementary rods to complementary bubbles as the volume fraction is 
decreased from 1 to 0. With the help of Eq.\ (\ref{fd}) one finds that the four critical values of $\chi$ are given by $G(1-\chi)=H(1-\chi)=H(\chi)=G(\chi)=0$, where 
\begin{subequations}
\bea
G(\chi)&\equiv& 23+4\chi-27\chi^{1/3}+5\ln\chi \, , \\[2ex]
H(\chi)&\equiv& 2\chi^2-\chi-3\chi\ln\chi-1 \, . 
\eea
\end{subequations}
This results in the critical values 
\be \label{chicrit}
\chi\simeq 0.785, \; 0.645, \; 0.355, \; 0.215    \, .
\ee
These values are completely independent of the details of the underlying microscopic model. Only the translation from $\chi$ into $\mu_n$ and the actual value of the free energy cost (\ref{DeltaF}) depends on the microscopic physics. 
We show the result of the step-like approximation in comparison to the full result in Fig.\ \ref{fig:deltaOm} for two fixed values of the surface tension. 
(As the full calculation will show, the surface tension itself is a function of $\mu_n$.) 

\subsection{Numerical evaluation}
\label{sec:full}

A more complete and reliable result is obtained by solving numerically the coupled differential equations (\ref{dOm}) without further approximations. 
To get a first idea of the form of the solution it is useful to discuss the behavior in the center and at the edge of the unit cell. This behavior is found by linearizing Eqs.\ (\ref{dOm})
about the boundary values. A detailed derivation, presented in appendix \ref{app:boundary}, shows that the lowest-order 
behavior is quadratic for all condensates and the electron chemical potential,
\begin{subequations} \label{square}
\bea
\bar{\sigma}(x) &\simeq& \bar{\sigma}_0+\left(\frac{\partial {\cal U}}{\partial\bar{\sigma}} + g_\sigma n_s\right)_0 \frac{(x-x_0)^2}{k} \, ,\\[2ex]
\bar{\omega}(x) &\simeq& \bar{\omega}_0+\left(m_v^2\bar{\omega} - g_\omega n_B\right)_0 \frac{(x-x_0)^2}{k} \, ,\\[2ex]
\bar{\rho}(x) &\simeq& \bar{\rho}_0+\left(m_v^2\bar{\rho} - g_\rho n_I\right)_0 \frac{(x-x_0)^2}{k} \, ,\\[2ex]
\mu_e(x) &\simeq& \mu_{e0}-e^2q_0 \frac{(x-x_0)^2}{k} \, .
\eea
\end{subequations}
Here, I have denoted the relevant coordinate collectively by $x$ for all three geometries, i.e., $x$ is a Cartesian coordinate for slabs, the cylindrical radial coordinate for rods, and the spherical radial coordinate for bubbles. The point at the center or the edge of the unit cell is denoted by $x_0$, and  $\bar{\sigma}_0, \bar{\omega}_0, \bar{\rho}_0, \mu_{e0}$ are the boundary values. The subscripts 0 at the coefficients of the quadratic terms indicate that the expressions are evaluated at the boundary. Moreover, $k=2$ at the outer boundary of the unit cell for all geometries, while in the center $k=2d$. Importantly, the boundary values themselves are not known a priori, and they cannot be computed by a local analysis. They rather have to be computed dynamically by 
solving the coupled differential equations over the entire domain of the unit cell. The coefficients of the quadratic terms are nothing but the right-hand sides of the Euler-Lagrange equations (\ref{dOm}). This shows that the profiles would be flat (the coefficient of the quadratic term would be zero) if the boundary values fulfilled the stationarity and neutrality equations.
The more a certain unit cell deviates from these properties at its boundaries, the less flat the profiles become. Besides these general insights into the solution, Eqs.\ (\ref{square}) are also useful as a check for the numerical solution, at least in the vicinity of the boundaries. 

I perform the full numerical calculation along the following steps: 

\begin{enumerate}

\item Fix a neutron chemical potential $\mu_n$ for which the mixed phase is preferred in the absence of Coulomb energy and surface tension, i.e., some $\mu_n$ in Fig.\ \ref{fig:omega} in the range of the (red) NQ curve.
At this given $\mu_n$, two charged phases with the same $\mu_e$ have the same pressure and form a globally neutral system with a volume fraction 
$\chi$. For notational compactness, let 
$\vec{v} \equiv (\bar{\sigma}, \bar{\omega}, \bar{\rho},\mu_e)$ and let $\vec{v}_{1}$ and $\vec{v}_2$ be the values of $\vec{v}$ in 
these two charged phases. 

\item Pick a geometry, i.e., bubbles, rods, or slabs. For bubbles and rods, decide which of the two phases occupies the interior of the cell. This yields 5 cases, and in all cases the Wigner-Seitz approximation ensures that the 
problem is effectively 
one-dimensional. 

\item Choose a domain, $x\in [x_{01}, x_{02}]$, such that $L=|x_{01}- x_{02}|$
is the width of the unit cell. In the case of bubbles and rods $x$ is a radial coordinate and thus $x_{01}=0$. Construct 4 initial guess functions  $\vec{v}(x)$, which are as close as possible to the (yet unknown) solutions. This can be done by interpolating between the values of the  homogeneous phases, $\vec{v}(x_{01}) = \vec{v}_1$, $\vec{v}(x_{02}) = \vec{v}_2$, for instance with the
tanh function in the case of the condensates and with the constant function $\mu_e(x)=\mu_e$ in the case of the electron chemical potential. 

\item Solve the 4 Euler-Lagrange equations (\ref{dOm}) for $\vec{v}$ numerically with the successive over-relaxation method\footnote{The 
same method was used in the calculation 
of the surface tension of the VQ, VN, NQ transitions for symmetric nuclear matter \cite{Fraga:2018cvr} and for flux tube profiles of multi-component 
superconductors \cite{Haber:2017kth,Haber:2017oqb}. I am working with a grid of about $5000$ points on the domain, corresponding to a lattice spacing of about $10^{-3}\, {\rm fm}$, and with up to about $10^6$ iterations steps. These numbers were obtained by checking convergence for selected cases to reach an accuracy such that the results of all plots in the paper would be basically indistinguishable if more grid points or iterations were used. In the slab geometry convergence is much quicker and a much coarser lattice and fewer iterations lead to the same results.}. Since the initial guess functions contain a nontrivial electric charge distribution, the Poisson equation induces a nontrivial behavior $\mu_e(x)$. In particular, $\mu_e(x)$ will differ at the boundaries from the (constant) starting value. This, in turn, 
also changes the condensates at the boundaries. Therefore, it is important to keep all boundary values dynamical in the relaxation procedure. 

\item Enforce electric neutrality in each iteration step of the relaxation procedure: By integrating the charge density over the unit cell, the total
charge is a functional of the profiles computed at a given iteration step, in particular of the electron chemical potential $Q=Q[\mu_e(x)]$. Electric neutrality is enforced by adding a constant contribution $\varphi$ that solves the algebraic equation $Q[\mu_e(x)+\varphi]=0$. This defines the corrected electron chemical potential $\mu_e(x)+\varphi$ used for the next iteration step. After sufficiently many iterations, the result is a stable interface for a given geometry and given $L$, where the boundary behavior of $\vec{v}(x)$ is given by Eqs.\ (\ref{square}).

\item Repeat the procedure for different $L$. For instance, if $L$ was chosen 
to be smaller than the (yet unknown) favorable $L$, successively increase $x_{02}$ by a small amount (while keeping $x_{01}$ fixed). It is useful to extend the solution from the previous, smaller, domain to the larger domain in order to use it as an initial guess, for instance by adding constant pieces to the profiles. Increasing the domain leads to nontrivial changes, including small changes in the boundary values, and therefore the relaxation procedure has to be repeated for every $L$. 

\item For each $L$, compute the free energy per unit volume $F$ via 
Eq.\ (\ref{F}) and determine the (local) minimum of $F$ as a function of $L$ if such a minimum exists.

\end{enumerate}

The final result, i.e., the preferred size of the unit cell together with the profile functions, only depends on $\mu_n$ and the chosen geometry. Quantities such as $L_0$, $\chi$, and $\Sigma$, needed to construct the step-like approximation, play no role in the calculation, everything is determined dynamically.     

\subsection{Results and discussion}
\label{sec:results}

\begin{figure} [t]
\begin{center}
\includegraphics[width=\columnwidth]{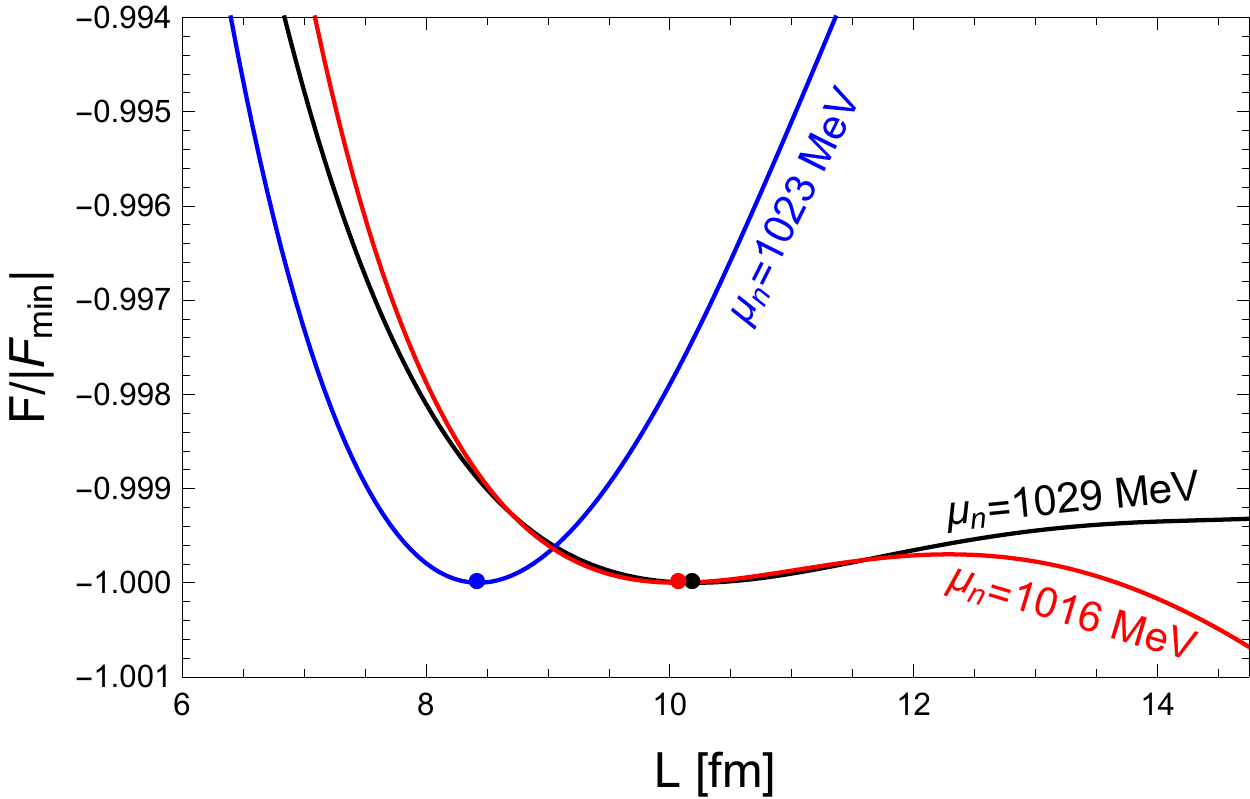}

\includegraphics[width=\columnwidth]{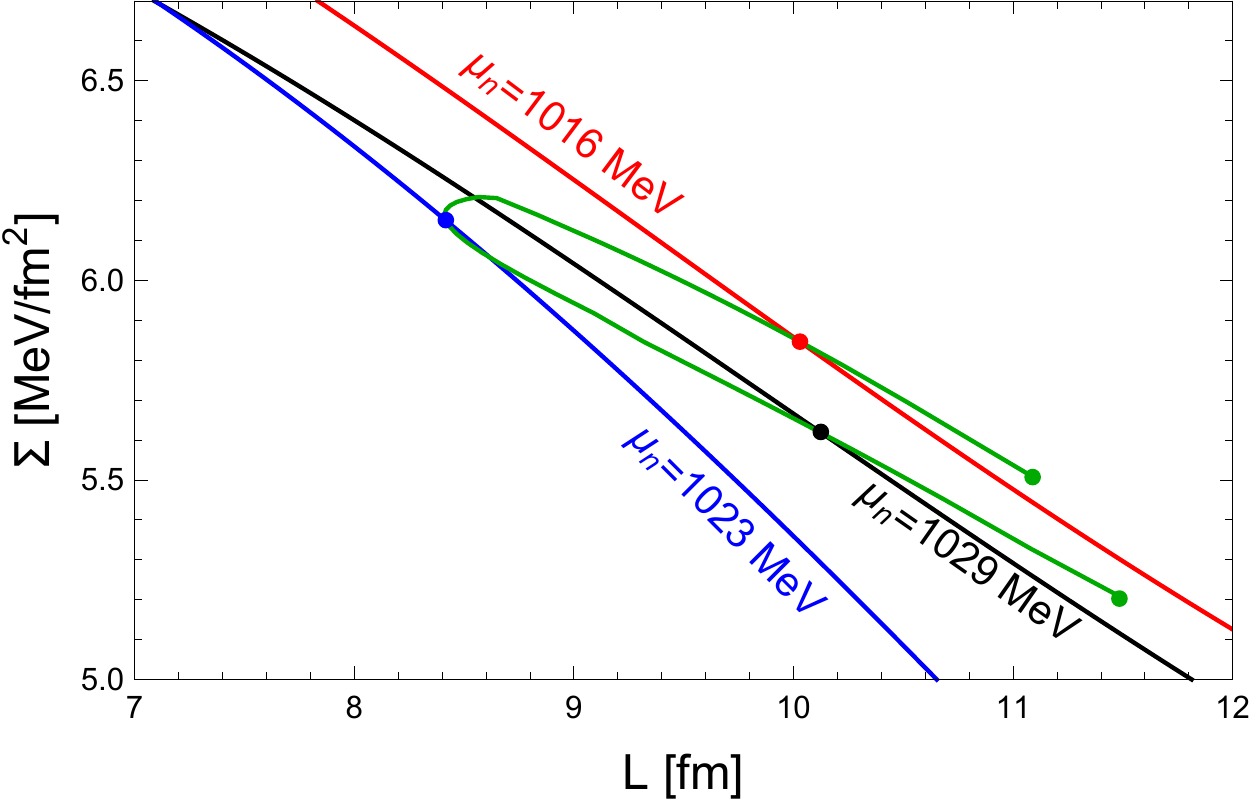}
\caption{{\it Upper panel:} Free energy per unit volume $F$ for slabs normalized to its value at the (local) minimum $F_{\rm min}$ for three different neutron chemical potentials as a function of the width of the unit cell $L$.  {\it Lower panel:} Surface tension of the slab configuration for the same neutron chemical potentials. The green line is the surface tension at the minimum for all $\mu_n$ where a minimum exists. Here and in all following results, $K=252\, {\rm MeV}$. 
}
\label{fig:Sigma}
\end{center}
\end{figure}

Using the slab geometry as an example, I will first explain some details of the results and afterwards present the main conclusions, bringing together the results of all three geometries.

\subsubsection{Unit cells and profiles for slabs}
\label{sec:results1}

\begin{figure*} [t]
\begin{center}
\includegraphics[width=\columnwidth]{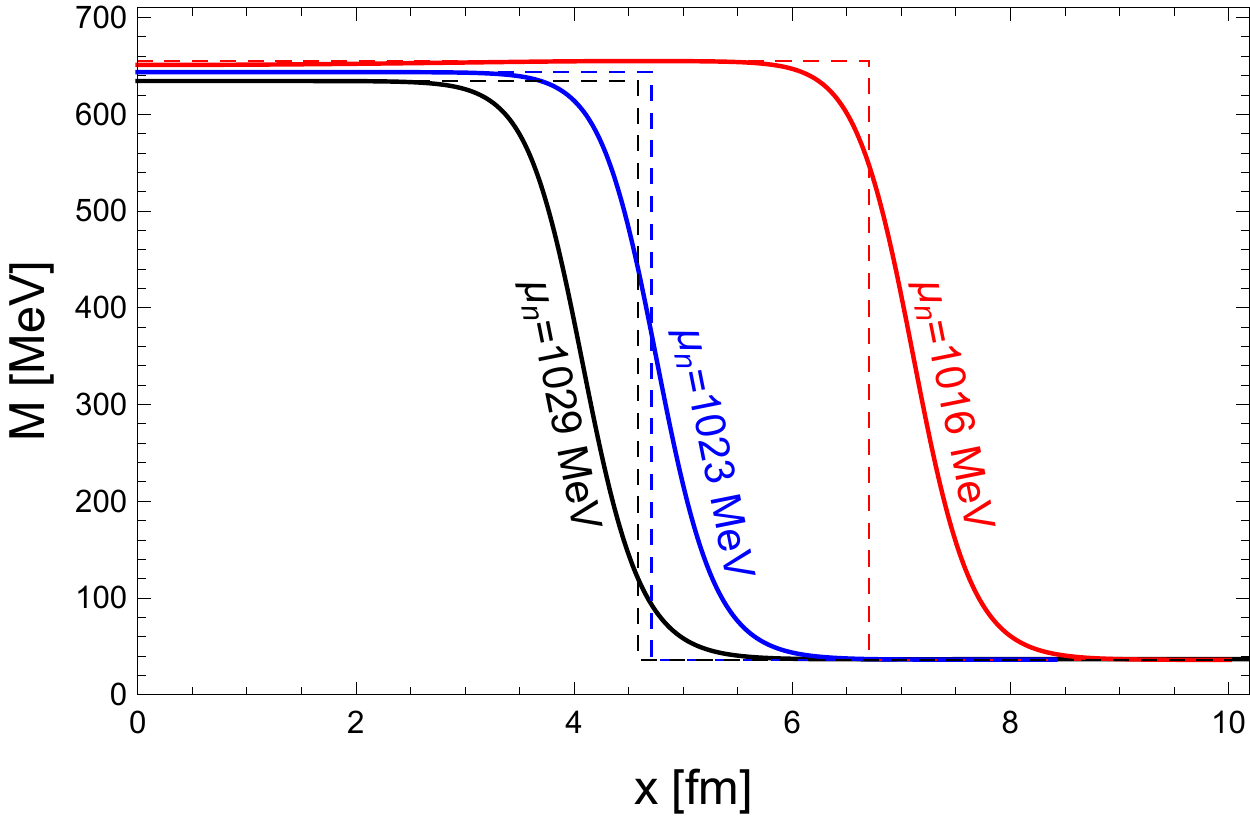}
\includegraphics[width=\columnwidth]{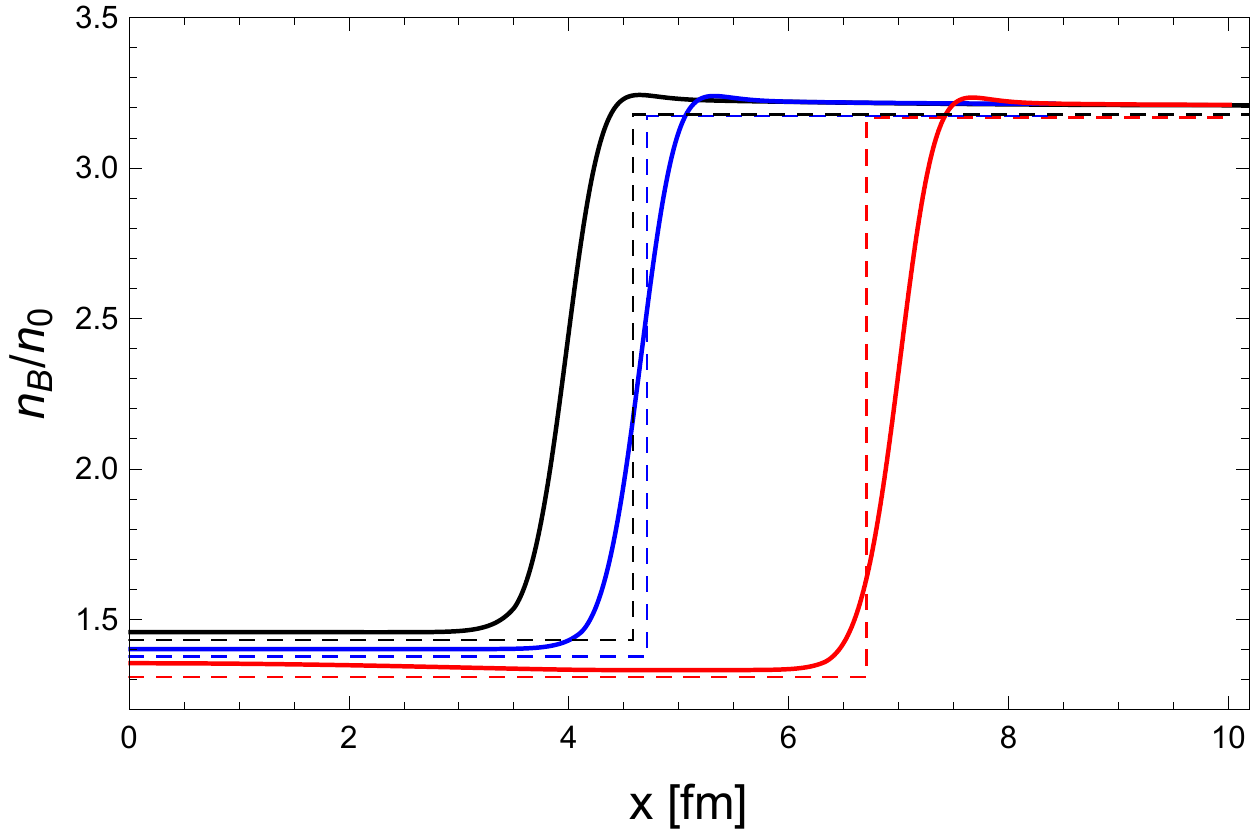}

\includegraphics[width=\columnwidth]{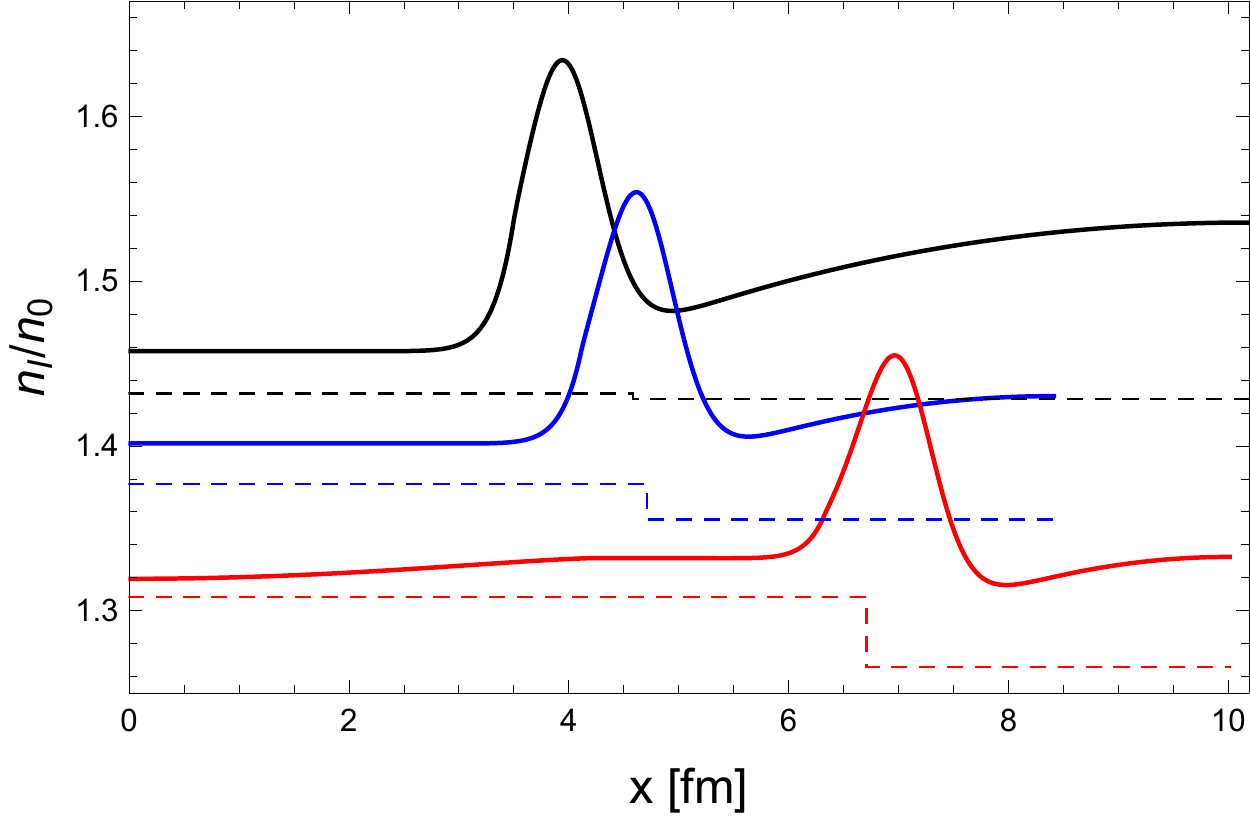}
\includegraphics[width=\columnwidth]{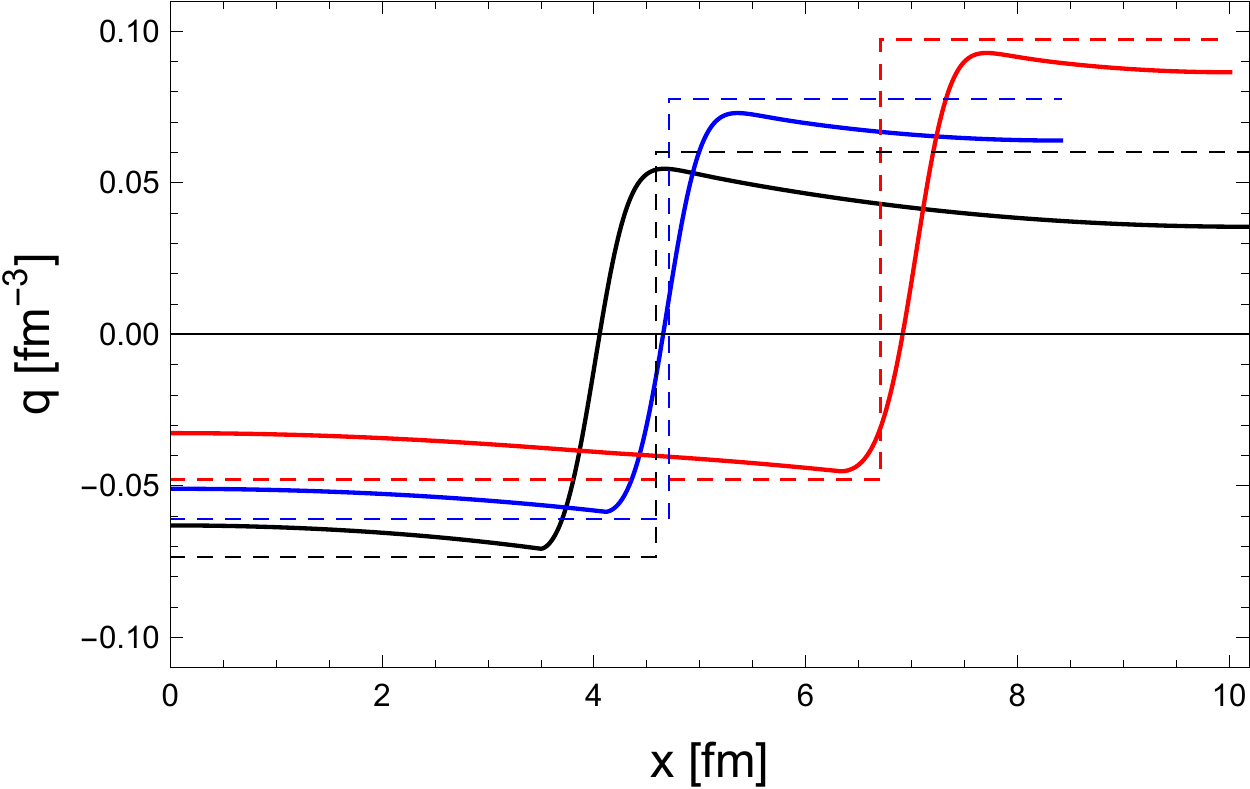}
\caption{Slab profiles of the effective nucleon mass $M$,  baryon number density $n_B$, isospin number density $n_I$, and charge density $q$ for the same three neutron chemical potentials as in Fig.\ \ref{fig:Sigma} at the energetically preferred widths of the 
Wigner-Seitz cell $L\simeq10.0\,{\rm fm}$ (red), $8.41\,{\rm fm}$ (blue), $10.2\, {\rm fm}$ (black). The chirally broken (restored) phase resides at the left (right) end of the scale in each plot. The thin dashed lines are the values obtained ignoring surface tension and Coulomb energy with a step-like profile and volume fractions of the chirally broken phase $\chi=0.67$ (red), $0.56$ (blue), $0.45$ (black).  
}
\label{fig:profiles}
\end{center}
\end{figure*}

In Fig.\ \ref{fig:Sigma}, results for three different neutron chemical potentials
$\mu_n$ are shown as a function of the width of the unit cell $L$. In the upper panel the free 
energy, calculated from Eq.\ (\ref{F}), is plotted. One of the curves has a very pronounced minimum, as one would expect from the step-like approximation. However, moving to smaller and larger chemical potentials, i.e., 
toward more imbalanced volume fractions, the minimum becomes more shallow. For the (red and black) curves shown here the minimum barely exists and indeed the minimum ceases to exist for slightly smaller (red) or larger (black) chemical potentials. At least for the curve $\mu_n= 1016\, {\rm MeV}$, it is obvious that the mixed phase with the slab geometry is metastable, i.e., the minimum is only local, not global. The system seems to be able to reduce its free energy by forming larger and larger unit cells. The corresponding profiles show that the volume partition between the two phases becomes more and more extreme, i.e., the tendency to form unit cells of large $L$ is nothing but the tendency to create uniform matter. Since the numerics become challenging for very large $L$, it is  difficult to 
continue the curve much further and show that it indeed asymptotes to the free energy of the uniform phase. But, it is easy to check that for the two cases $\mu_n= 1016\, {\rm MeV}$ and $1029\, {\rm MeV}$ the 
free energy of the uniform state (N in the case of the smaller $\mu_n$, Q for the larger one) is indeed smaller than the free energy of the local minimum, as shown explicitly in Fig.\ \ref{fig:deltaOm} below. 

The lower panel of Fig.\ \ref{fig:Sigma} shows the surface tension, computed from 
Eq.\ (\ref{Sigma}). First of all, 
not surprisingly, one sees that the surface tension is not a constant. For instance, it decreases as the unit cell gets larger for fixed $\mu_n$. The figure also shows the values of the surface tension at the energetic minimum for all $\mu_n$  for which the slab solution is at least a local minimum. For these values ones reads off $\Sigma \simeq (5.2 - 6.2)\, {\rm MeV}/{\rm fm}^2$. These values are somewhat smaller than but very similar to the 
surface tension computed in the same model with isospin-symmetric matter from a domain wall configuration with semi-infinite phases at the chiral phase transition \cite{Fraga:2018cvr}\footnote{With the chosen parameters $M_0=0.75\, m_N$, $K = 252\, {\rm MeV}$, symmetric nuclear matter is only metastable in the given model, such that in Ref.\ \cite{Fraga:2018cvr} only the surface tension of the transition between the vacuum and the chirally restored phase was calculated at this parameter point, see for instance Fig.\ 6 in this reference. The same figure shows that for slightly larger values of $K$ or $M_0$, where symmetric nuclear matter {\it is} stable, one finds $\Sigma \simeq 8\, {\rm MeV}/{\rm fm}^2$, not far from the result obtained for asymmetric matter here.}.

Figure \ref{fig:profiles} shows the corresponding profiles across one wall of the slab for the same three neutron chemical potentials as in Fig.\ \ref{fig:Sigma}. In each case, the profiles are shown for the $L$ at the local minimum. The coordinates are chosen such that $x=0$ corresponds to the center of the slab in all three cases, while the right end of the scale corresponds to the size of the largest of the three unit cells. For comparison, the figure also shows the constant values obtained by ignoring surface tension and Coulomb effects, i.e., the values that give the results for the NQ phase in Figs.\ \ref{fig:omega} and \ref{fig:MMue}. The plot confirms that the boundary values of the profiles deviate -- sometimes significantly -- from these constant values. The step that connects the constant values is chosen to reproduce the volume fractions from the NQ phases of Figs.\ \ref{fig:omega} and \ref{fig:MMue}, using the preferred unit cell size from the full calculation\footnote{The step-like approximation has, by construction, the same volume fraction, but a different 
preferred unit cell size, i.e., the dashed curves do not exactly represent the step-like approximation from Sec.\ \ref{sec:step}.}. One can see from the curves for the largest and the smallest $\mu_n$ that the full calculation tends to 
create a more pronounced imbalance between the two phases than predicted from the
step-like curves. 

The profiles for the effective mass $M$ and the baryon number $n_B$ illustrate 
that the wall  interpolates between the chirally broken (large $M$), low-density phase and the chirally restored (small $M$), high-density phase. One characteristic feature of the slab is the accumulation of isospin density at the interface, as the lower left panel shows. The profiles of the electric charge density $q$ demonstrate the screening effect, i.e., the accumulation of negative and positive charge carriers close to the interface. This effect reduces the Coulomb energy cost and thus works in favor of large unit cells. One might wonder why the charge accumulation is not an even more pronounced effect. But of course one has to keep in mind that reducing the Coulomb energy is not the only effect at work. 
A more extreme charge accumulation is for instance prevented by the tendency of the condensates to remain as close as possible to the local minimum they would assume in the absence of electrostatic energy costs, in order to reduce the potential energy of the configuration.

The charge density profiles also show a cusp close to the interface at negative $q$. This cusp arises because there is a regime where there are no protons in the chirally broken phase: for all $x$ up to the cusp for $\mu_n=1023\, {\rm MeV}$ and $1029 \, {\rm MeV}$
(blue and black curves) and for $x\simeq 4 \, {\rm fm}$ up to the cusp for $\mu_n=1016\, {\rm MeV}$ (red curve). Thus, in this regime, the system decides to create baryon number in the chirally broken phase only from neutrons, with the lepton gas providing negative charge. This is possible because the chirally restored phase 
is positively charged. At this point, it is useful to recall that although in some sense the chirally restored phase can be interpreted as quark matter there are important differences to realistic quark matter. First of all, this phase can, at best, be identified with two-flavor quark matter because strangeness is not included in the model. A priori, there is nothing wrong with this restriction because it is plausible that in the transition region under realistic neutron star conditions the strange quarks are sufficiently heavy to only play a very minor role \cite{Fujimoto:2019sxg}. However, in the context of locally charged mixed phases it is important to note that the chirally restored phase in the present model has different charge carriers than realistic two-flavor quark matter, which is made of up and down quarks with charges $2/3\, e$ and $-1/3\, e$. Most notably, in the present model there is no negative charge carrier that also carries baryon number, i.e., no analogue of the down quark. All negative charge is carried by the leptons. This has to be kept in mind in the interpretation of the results. In fact, this shortcoming of the model provides a strong motivation to include strangeness in future studies. The reason is that, even if hyperons might not play any role in the chirally broken phase, there will be negative baryonic charge carriers in the system, which, in the chirally restored phase, may provide a more realistic description of actual quark matter.

\begin{figure} [t]
\begin{center}
\includegraphics[width=\columnwidth]{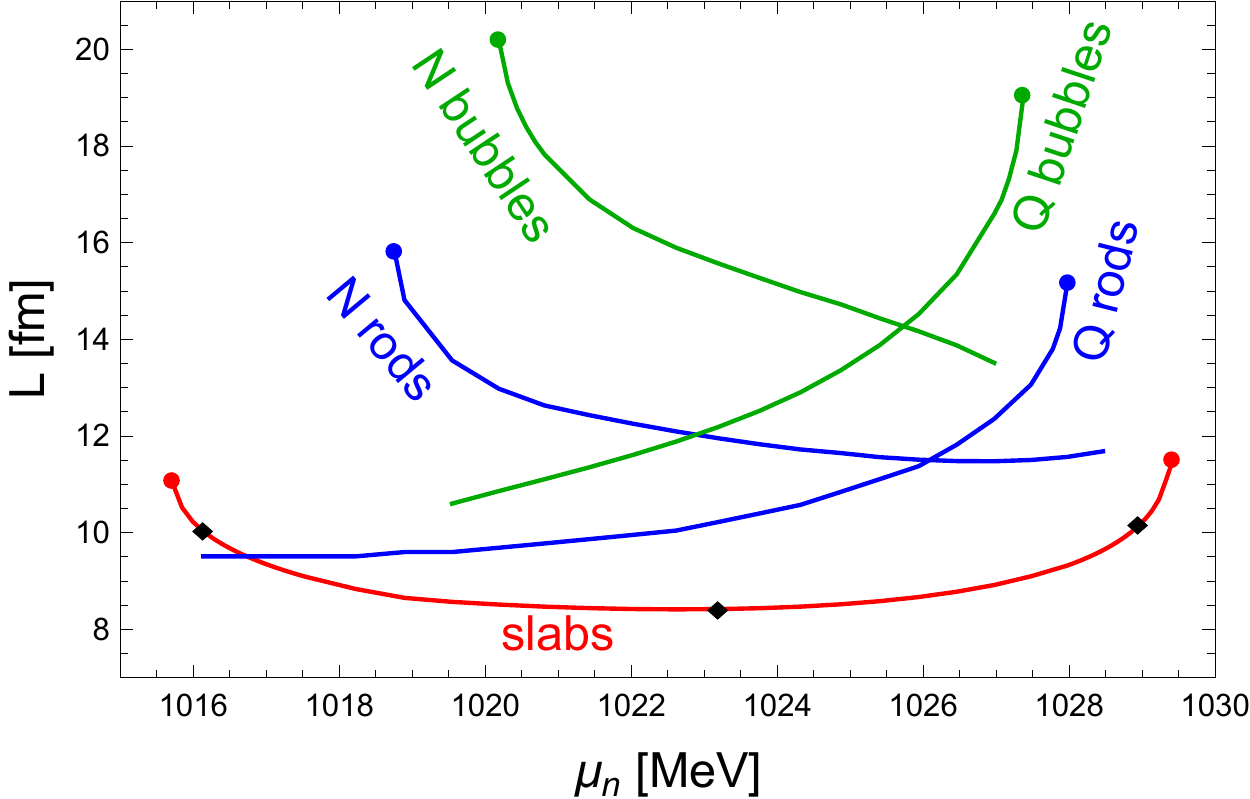}
\caption{Energetically preferred widths $L$ of the unit cells  as a function of the neutron chemical potential $\mu_n$ for bubbles (green), rods (blue), and slabs (red). Spherical dots mark the end of the line beyond which there is no stable mixed phase solution; at the other end of the curves for bubbles and rods the numerics start becoming too difficult. The (black) diamonds indicate the three points used for Figs.\ \ref{fig:Sigma} and \ref{fig:profiles}.  
}
\label{fig:LWS}
\end{center}
\end{figure}

\begin{figure} [t]
\begin{center}
\includegraphics[width=\columnwidth]{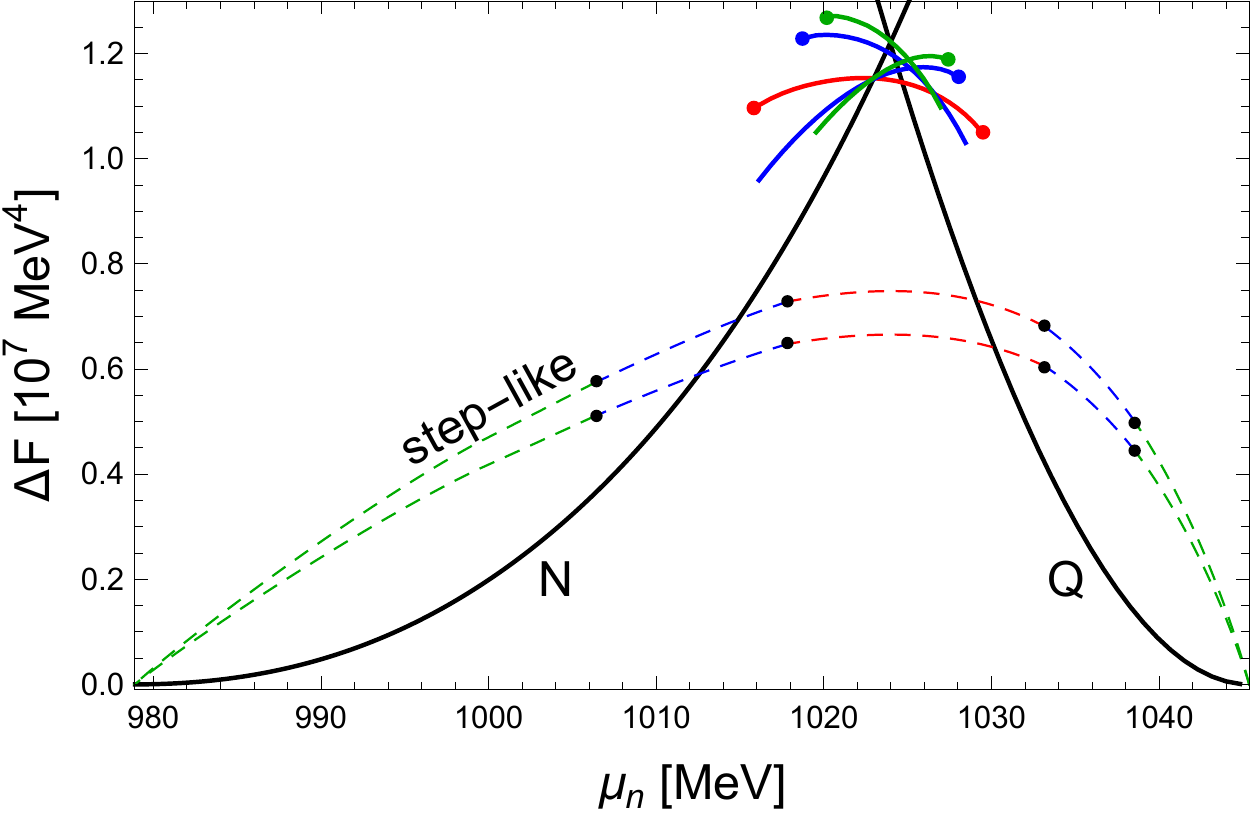}

\includegraphics[width=\columnwidth]{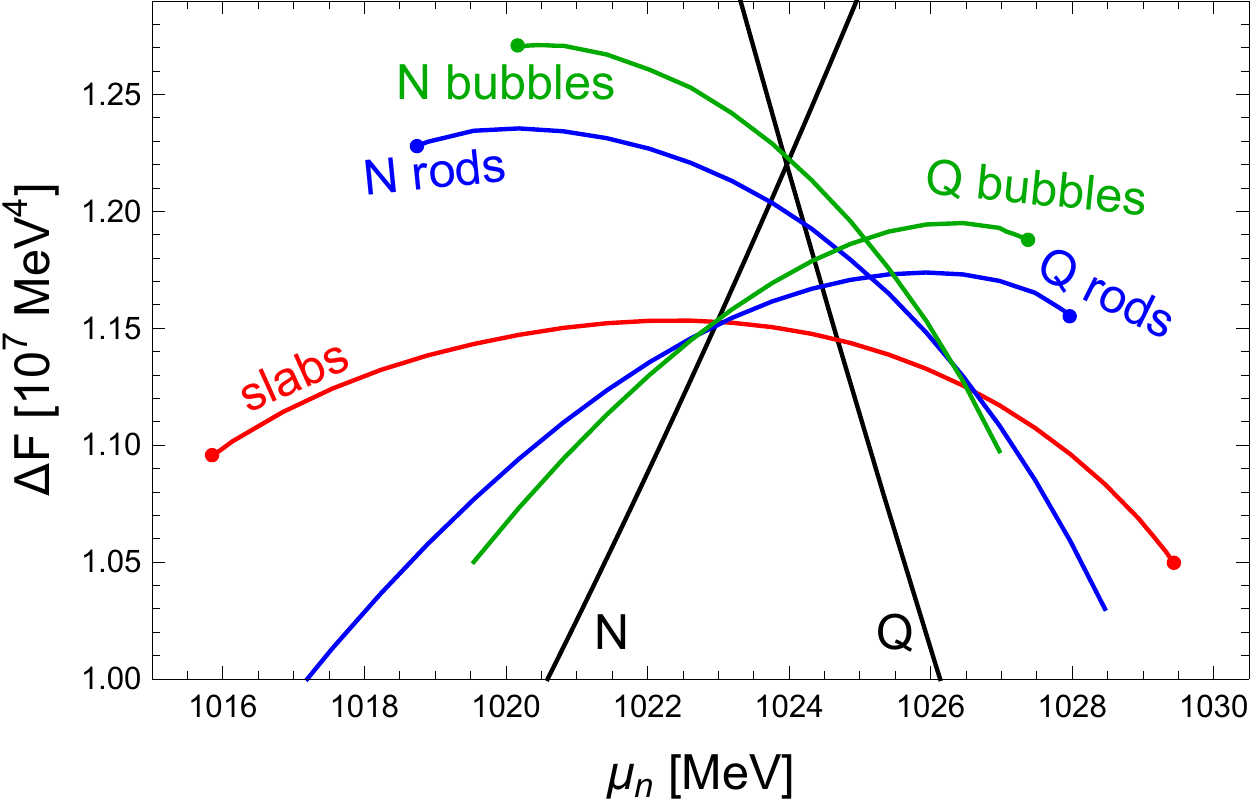}
\caption{{\it Upper panel:} Free energies per unit volume of homogeneous 
phases (black), chiral pasta phases from the full numerical calculation (solid red, green and blue curves) and from the step-like approximation using the 
two constant surface tensions $\Sigma = 5.2\, {\rm MeV}/{\rm fm}^2$ and $6.2\, {\rm MeV}/{\rm fm}^2$  (thin dashed curves). All free energies are relative to the mixed phase without any Coulomb effects and surface tension, i.e., relative to the (red) NQ curve in Fig.\ \ref{fig:omega}. {\it Lower panel:} Magnification of the region where chiral pasta is stable or metastable, showing that slabs (and possibly Q rods in a tiny regime) are the preferred pasta configuration. 
}
\label{fig:deltaOm}
\end{center}
\end{figure}

\subsubsection{Comparison of all geometries}
\label{sec:results2}

Figures \ref{fig:LWS} and \ref{fig:deltaOm} combine the results for all geometric structures considered in this paper. For the bubbles and rods one has to distinguish whether nuclear matter forms bubbles or rods immersed in quark matter ('N bubbles', 'N rods') or vice versa ('Q bubbles', 'Q rods'). Fig.\ \ref{fig:LWS} shows the sizes $L$ of the unit cell that minimize the free 
energy for a given neutron chemical potential $\mu_n$, i.e., this is the 
result of the analysis illustrated by the upper panel of Fig.\ \ref{fig:Sigma}, now applied to all five geometries. In the case of the slabs, the three chemical potentials used for Figs.\ \ref{fig:Sigma} and \ref{fig:profiles} are marked
by diamonds.
The curve for the slabs ends on both sides at a point, marked by a dot, beyond which the local minimum of the 
mixed phase configuration ceases to exist. In the case of bubbles and rods the same disappearance of the local minimum occurs on one end of the curves. This is the end where the preferred uniform phase is the one that fills the bubbles or rods. 
In other words, these endpoints indicate the critical $\mu_n$'s {\it above} which Q bubbles and Q rods are no longer a minimum and the critical $\mu_n$'s {\it below} which N bubbles and N rods are no longer a minimum. The other end of the curves is in the regime where by reducing the size of the bubbles or rods to zero one arrives at the energetically preferred uniform phase. On this side, beyond a certain point the 
numerical relaxation algorithm does not yield any non-trivial profiles, the profiles rather relax to the homogeneous solution. This may well be a purely numerical problem, I cannot exclude that non-trivial profiles continue to exist beyond that point. Therefore, in Figs.\ \ref{fig:LWS} and \ref{fig:deltaOm} I have not marked these ends of the lines by a dot, suggesting that the lines potentially continue. Nevertheless, the range where I did find non-trivial solutions is sufficient in the sense that at the point where I have to stop, one of the uniform phases is already favored, i.e., this numerical problem occurs in the regime where the pasta phases are metastable anyway. 

Figure \ref{fig:deltaOm} shows the main result of the paper by comparing the free energies of the uniform phases with the ones of the mixed phases from the full numerical calculation and from the step-like approximation. All free energies 
are plotted relative to the free energy of the mixed phase without any surface and Coulomb effects. This free energy difference is denoted by $\Delta F$, i.e., 
$\Delta F=0$ corresponds to the (red) NQ curve in Fig.\ \ref{fig:omega}. The step-like approximation shows the sequence of phases Q bubbles $\to$ Q rods $\to$ slabs $\to$ N rods $\to$ N bubbles, separated by black dots at the points given by the critical volume fractions (\ref{chicrit}). This sequence is generic for this approximation and does not depend on the details of the system.
As discussed above, the step-like approximation
requires a value for the surface tension as an input. For a 
rough comparison with the full result I have chosen the two constant values $\Sigma=5.2\, {\rm MeV}/{\rm fm}^2$ and $\Sigma=6.2\, {\rm MeV}/{\rm fm}^2$, motivated by the minimal and maximal values obtained from the numerical calculation, see lower panel of Fig.\ \ref{fig:Sigma}. A density dependent surface tension between these two boundaries would then yield a curve within the band given by the two curves of constant surface tension. Considering only the most preferred configuration for each $\mu_n$, the step-like approximation with this choice of the surface tension predicts a 
sequence of phase transitions from uniform nuclear matter at low densities to Q rods, then to slabs, and then to uniform quark matter at high densities.
I have checked that surface tensions $\Sigma \gtrsim 13\, {\rm MeV}/{\rm fm}^2$ are needed to lift the entire dashed curve above the free energies of the uniform phases, i.e., to make the mixed phases disfavored for all $\mu_n$. 

The solid lines show the full result, with the lower panel giving a more detailed view of the chiral pasta regime. The results indicate that the mixed phases are less favorable than predicted by the step-like approximation with the same surface tension. They exist in a range of neutron chemical potentials of $\Delta \mu_n \simeq 2\, {\rm MeV}$, compared to 
$\Delta \mu_n \simeq 15 \, {\rm MeV}$ for the step-like approximation and $\Delta\mu_n\simeq 66 \, {\rm MeV}$ if Coulomb and surface effects are entirely neglected. Interestingly, the preferred structures seem to be identical to the prediction of the step-like approximation: also in the full result most of the mixed phase regime consists of slabs, possibly with a small region of Q rods at the low-density end. All transitions between the different homogeneous and inhomogeneous structures are of first order, and in a more complete treatment more complicated geometries, or mixtures of different pasta structures, might occur in the vicinities of these first-order transitions.   

One might wonder whether it is a coincidence that the chiral pasta is just barely preferred and that predominantly slabs, not bubbles or rods, seem to appear. Is it possible that pasta is never preferred, i.e., that it is metastable for all densities? 
One way to check this would be to repeat the calculation for different model parameters. After all, there is some freedom in choosing the parameters due to the uncertainty in the saturation properties $M_0$, $S$, and $K$. Due to the relatively tedious numerical calculation it is not easy to present a systematic 
survey of the parameter space, and I have not repeated the full-fledged calculation for different parameters. I did, however, repeat the calculation of the slabs 
for a larger value  of the incompressibility at saturation $K$  (still within the experimentally predicted range) while keeping $M_0$ and $S$ fixed and found the free energy to be larger for any $\mu_n$ than that of the homogeneous phases. This 
suggests that there is no pasta phase at all for large $K$. This is perhaps expected  since it was already demonstrated 
in Fig.\ \ref{fig:mixed1} that the potential range for a mixed phase, even in the absence of surface tension and Coulomb energy, becomes smaller as $K$ is increased. Therefore, in this regime, the energy costs from surface and Coulomb effects can easily  destroy the mixed phases. Consequently, the 
quantitative and even some of the qualitative details of Fig.\ \ref{fig:deltaOm} should not be taken too literally. Even within the chosen model they depend on the exact values of the model parameters.

\section{Summary and outlook}
\label{sec:summary}

I have computed pasta structures and their free energies consistently within a single model at the chiral phase transition. To this end, I have employed a phenomenological model where nucleons and mesons are the fundamental degrees of freedom,
to which non-interacting electrons and muons are added.  For model parameters that reproduce saturation properties of nuclear matter and with the constraints of electric charge neutrality and beta equilibrium, this model shows a first-order chiral phase transition at zero temperature and in the absence of mixed phases. I have first identified the region in the vicinity of this phase transition where globally, but not locally, neutral mixed phases are possible without taking into account surface and Coulomb effects. These effects have then been taken into account by calculating the  
profiles of the meson condensates and the electrostatic potential in a  
consistent way, solving the Euler-Lagrange equations for the condensates coupled with the Poisson equation for the electrostatic potential. Doing this for 
three different geometries -- bubbles, rods, and slabs -- and determining the preferred sizes of the unit cells dynamically for each case, this calculation yields the free energies of the different pasta structures. I have compared these free energies with each other, with the free energy of the homogeneous phases, and with the free energy obtained from a simple step-like approximation of the profiles, which is often used in the literature. As a result, I have found that for the chosen model parameters chiral pasta is favored in a vicinity of the first-order phase transition which is only about 2 MeV wide,  measured in the neutron chemical potential, and that the predominant structure that appears consists of slabs. As a by-product, I have extracted the density dependent surface tension, computed from domain walls in the slab geometry, at the energetically preferred unit cell sizes. I have found values 
$\Sigma \simeq (5.2 - 6.2) \, {\rm MeV}/{\rm fm }^2$, which are comparable to, although slightly smaller than the surface tension of isospin-symmetric nuclear matter at the chiral phase transition in the same model. 
All these results depend on the model parameters, which are not uniquely fixed due to the 
experimental uncertainty of some of the saturation properties of nuclear matter. 
Exploring this parameter space and the fate of the mixed phases systematically is left for future work, but I have pointed out that larger values of the incompressibility at saturation lead to even smaller and eventually vanishing regimes for the pasta phases. 

I have used various approximations, which can be improved in future studies. For instance, one may go beyond the mean-field approximation, which becomes particularly relevant for nonzero temperatures, or improve on the Thomas-Fermi and Wigner-Seitz approximations that I have used for the inhomogeneous phases. Also, it should be emphasized that the high-density phase in the present model is only a very crude approximation to a real-world quark matter phase. 
In particular, there are no baryonic charge carriers with negative electric charge, which is relevant for the structure of the mixed phases. Therefore, one interesting extension would be to include strangeness, which introduces negative charge carriers in the form of hyperons and their chirally restored counterparts. 
Another extension would be to allow for an inhomogeneous chiral condensate in the form of a chiral density wave, and study the interplay of the mixed phases with this inhomogeneous structure. Furthermore, the present model suggests that a vacuum-quark mixed phase might be adjacent in chemical potential to a quark-hadron mixed phase if the chiral phase transition is sufficiently close to the baryon onset. As a consequence, a layer containing nuclear matter within an inhomogeneous phase in quark stars or even the existence of a three-component mixed phase are conceivable. I have not addressed these intriguing possibilities in detail here, and it would be interesting to study them in the future.  
Finally it would of course be desirable to implement the results of this paper into a calculation of the structure of a neutron star or into a simulation of a neutron star merger, which potentially probes the fate of chiral pasta at nonzero temperatures.

\begin{acknowledgments}
I am grateful to M.\ Alford, S.\ Carignano, E.\ Fraga, M.\ Hippert, A.\ Pfaff, K.\ Rajagopal, and A.\ Sedrakian
for valuable discussions and comments. I am supported by the Science \& Technology Facilities Council (STFC) in the form of an Ernest Rutherford Fellowship.
\end{acknowledgments}

\appendix

\section{Asymmetry energy and incompressibility at saturation}
\label{app:sym}

The asymmetry energy is defined as \cite{glendenningbook}
\be \label{Sdef}
S = \frac{1}{2} \frac{\partial^2 (\epsilon/n_B)}{\partial (n_I/n_B)^2} = \frac{n_B}{2} \frac{\partial^2 \epsilon}{\partial n_I^2} = \frac{n_B}{2} \frac{\partial\mu_I}{\partial n_I} \, , 
\ee
where $\epsilon$ is the energy density, where the derivatives are taken at fixed $n_B$, and where the thermodynamic relation 
\be
\mu_I = \frac{\partial\epsilon}{\partial n_I} 
\ee
has been employed. The isospin chemical potential 
is $\mu_I=(\mu_n-\mu_p)/2$ and thus with Eqs.\ (\ref{mustar}), (\ref{eqrho}), and (\ref{nnnp}) one computes 
\bea \label{muI}
\mu_I 
&=& \frac{g_\rho^2}{m_v^2}n_I+\frac{1}{2}\left(\sqrt{\left(\frac{3\pi^2}{2}\right)^{2/3}(n_B+n_I)^{2/3}+M^2} \right. \non[2ex]
&&\left. - \sqrt{\left(\frac{3\pi^2}{2}\right)^{2/3}(n_B-n_I)^{2/3}+M^2}\right) \, .
\eea
Consequently, 
\bea
&&\frac{\partial\mu_I}{\partial n_I} = \frac{g_\rho^2}{m_v^2} +\frac{1}{6}\left(\frac{3\pi^2}{2}\right)^{2/3}\left[\frac{1}{\mu_n^*(n_B+n_I)^{1/3}}\right. \non[2ex]
&&\left. +\frac{1}{\mu_p^*(n_B-n_I)^{1/3}}\right]
+\frac{M}{2}\frac{\partial M}{\partial n_I}\left(\frac{1}{\mu_n^*}-\frac{1}{\mu_p^*}\right) \, .
\eea
For the evaluation of this expression at the symmetric point, where 
$\mu_n^*=\mu_p^*$, the derivative of $M$ with respect to $n_I$ is not needed. With $n_I=0$, evaluating the result at saturation and inserting it into the definition (\ref{Sdef}) yields
\be
S = \frac{k_F^3}{3\pi^2}\frac{g_\rho^2}{m_v^2}+\frac{k_F^2}{6\mu_B^*} \,, 
\ee
where 
\be
\mu_B^*=\sqrt{k_{F}^2+M^2} \, .
\ee
 This is in accordance with Eq.\ (4.218) in Ref.\ \cite{glendenningbook} (after $g_\rho\to g_\rho/2$ due to the different definition of $g_\rho$) and gives 
Eq.\ (\ref{grho}) in the main text.  

The derivation of the incompressibility is similar. Since the incompressibility is needed for symmetric matter at saturation and no isospin derivatives are required, one can perform the entire calculation in the symmetric scenario. The incompressibility is defined as
\be\label{defK}
K = 9 n_B \frac{\partial^2\epsilon}{\partial n_B^2} = 9 n_B\frac{\partial \mu_B}{\partial n_B} \, , 
\ee
again using the thermodynamic relation
\be
\mu_B = \frac{\partial\epsilon}{\partial n_B} \, . 
\ee
In analogy to Eq.\ (\ref{muI}) one has 
\be
\mu_B = \frac{g_\omega^2}{m_v^2}n_B+ \sqrt{\left(\frac{3\pi^2}{2}\right)^{2/3}n_B^{2/3}+M^2} \, , 
\ee
where the omega condensate has been eliminated with the help of Eq.\ (\ref{eqomega}). Consequently,
\be
\frac{\partial\mu_B}{\partial n_B} = \frac{g_\omega^2}{m_v^2} +\frac{1}{3}\left(\frac{3\pi^2}{2}\right)^{2/3}\frac{1}{\mu_B^*n_B^{1/3}}
+\frac{M}{\mu_B^*}\frac{\partial M}{\partial n_B} \, ,
\ee
and thus, with the definition (\ref{defK}),
\be\label{K1}
K = \frac{6k_F^3}{\pi^2}\frac{g_\omega^2}{m_v^2} + \frac{3k_F^2}{\mu_B^*} + \frac{9n_BM}{\mu_B^*}\frac{\partial M}{\partial n_B} \, .
\ee
Now the derivative of $M$ does not drop out, and one has to compute it with the help of Eq.\ (\ref{eom1}). First note that at zero temperature
\be \label{ns1}
n_s = \frac{2M}{\pi^2}\int_0^{k_F} dk\frac{k^2}{\epsilon_k} \, , \qquad \epsilon_k = \sqrt{k^2+M^2} \, .
\ee
Taking the derivative with respect to $M$ on both sides of Eq.\ (\ref{eom1}) 
yields
\be\label{dMdnB}
\frac{\partial M}{\partial n_B} = -\frac{\frac{\partial n_s}{\partial n_B}}{\frac{\partial^2{\cal U}}{\partial M^2}+\frac{\partial n_s}{\partial M}} \, .
\ee
With Eq.\ (\ref{ns1}) one finds
\begin{subequations}
\bea
\frac{\partial n_s}{\partial M} &=& \frac{2}{\pi^2}\int_0^{k_F} dk\frac{k^4}{\epsilon_k^3} \, , \\[2ex]
\frac{\partial n_s}{\partial n_B}&=& \frac{\partial n_s}{\partial k_F}\frac{\partial k_F}{\partial n_B} = \frac{M}{\mu_B^*} \, .
\eea
\end{subequations}
Inserting this into Eq.\ (\ref{dMdnB}) and the result into Eq.\ (\ref{K1}) yields the incompressibility
\bea\label{K}
K &=& \frac{6k_F^3}{\pi^2}\frac{g_\omega^2}{m_v^2} + \frac{3k_F^2}{\mu_B^*} \non[2ex]
&&- \frac{6k_F^3}{\pi^2}\left(\frac{M}{\mu_B^*}\right)^2\left[\frac{\partial^2{\cal U}}{\partial M^2}+\frac{2}{\pi^2}\int_0^{k_F} dk\frac{k^4}{\epsilon_k^3}\right]^{-1} \, , \hspace{0.8cm}
\eea
where the momentum integral can be done analytically,
\be 
\int_0^{k_F} dk\frac{k^4}{\epsilon_k^3} = \frac{k_F^3+3k_FM^2}{2\mu_B^*}-\frac{3M^2}{2}\ln\frac{k_F+\mu_B^*}{M} \, .
\ee

\section{Coulomb energy in step-like approximation}
\label{app:coulomb}

In order to compute the Coulomb energy density (\ref{WC}) in the step-like approximation one first needs to compute the electric field. Due to the symmetry of the problem in all three cases (bubbles, rods, and slabs in unit cells of corresponding shapes), this is best done with the method of Gaussian surfaces. The 
integral form of Gauss' law is (in Heaviside-Lorentz units)
\be \label{gauss}
\oint \vec{E}\cdot d\vec{S} = Q \, , 
\ee
where the integral is taken over a closed surface, which contains the total charge $Q$.
With the charge densities $\rho_1$ (inner phase, shaded in 
 Fig.\ \ref{fig:cells}) and $\rho_2$ (outer phase) and the volume fraction $\chi$ from Eq.\ (\ref{chi}), charge neutrality of the total 
cell can be written as 
\be\label{neutral}
\rho_1 - \rho_2 =\frac{\rho_1}{1-\chi} \, . 
\ee

\subsection{Bubbles}

For the spherical geometry one has $\vec{E}(\vec{r}) = E(r)\vec{e}_r$, where $r$ is the 
radial coordinate, and thus with a spherical Gaussian surface of radius $r$ Eq.\ (\ref{gauss}) yields
\be
E(r) = \frac{Q}{4\pi r^2} \, ,
\ee
where $Q$ is the charge contained in a sphere of radius $r$. 
The calculation of the electric field thus reduces to the calculation of that charge and one easily obtains 
\be
E(r) = \frac{1}{3} \left\{\begin{array}{cc} \rho_1 r & \mbox{for}\;\;0<r<L_0 \\[2ex]
\displaystyle{(\rho_1-\rho_2)\frac{L_0^3}{r^2}+\rho_2 r} & \mbox{for}\;\; L_0<r<L \end{array}\right. \, .
\ee
Piecewise integration over the radial coordinate then yields the Coulomb energy (\ref{WC}) per unit volume,
\bea \label{wCbubbble}
w_C
&=&\frac{\chi L_0^2}{30}\left[\rho_1^2(6-5\chi^{1/3})+\rho_2^2(9-5\chi^{1/3}-5\chi^{-2/3}\right.\non[2ex]
&&\left. +\chi^{-5/3})+5\rho_1\rho_2(2\chi^{1/3}-3+\chi^{-2/3})\right] \non[2ex]
&=& \frac{(\rho_1-\rho_2)^2L_0^2\chi}{10}(2+\chi-3\chi^{1/3}) \, ,
\eea
where the first result is general for any charge densities $\rho_1$, $\rho_2$, and in the second step the neutrality condition (\ref{neutral}) has been used. 

\subsection{Rods}

In this case, using cylindrical coordinates whose radial component I also denote by $r$, the electric field has the form $\vec{E}(\vec{r}) = E(r)\vec{e}_r$. Using a cylindrical Gaussian surface with length $L_z$ and radius $r$, Gauss' law  
gives 
\be
E(r) = \frac{Q}{2\pi r L_z} \, .
\ee
Calculating the enclosed electric charge $Q$ yields the electric field 
\be
E(r) = \frac{1}{2} \left\{\begin{array}{cc} \rho_1 r & \mbox{for}\;\;0<r<L_0 \\[2ex]
\displaystyle{(\rho_1-\rho_2)\frac{L_0^2}{r}+\rho_2 r} & \mbox{for}\;\; L_0<r<L \end{array}\right. \, .
\ee
Therefore, the Coulomb energy per volume is
\bea \label{wCrod}
w_C
&=&\frac{\chi L_0^2}{16}\left[\rho_1^2(1-2\ln\chi)+\rho_2^2(3-4\chi^{-1}\right.\non[2ex]
&&\left.+\chi^{-2}-2\ln\chi)+4\rho_1\rho_2(\ln\chi+\chi^{-1}-1)\right] \non[2ex]
&=& \frac{(\rho_1-\rho_2)^2L_0^2\chi}{8}(\chi-1-\ln\chi) \, ,
\eea
again using charge neutrality in the second step. 

\subsection{Slabs}

In this case, let the $x$ direction be perpendicular to the slab, i.e., the slab is extended in the $y$ and $z$ directions, and let the point $x=0$ be such that the $y$-$z$-plane is in the middle of the slab. Then one can focus on $x>0$ and the electric field is $\vec{E}(\vec{r}) = E(x)\vec{e}_x$ and $E(0)=0$ for symmetry reasons. As a Gaussian surface one can for instance take a cylinder extended in the $x$ direction with one end at $x=0$, such that the surface integral in 
Gauss' law only receives a contribution from the opposite end, 
\be
E(x) = \frac{Q}{A} \, ,
\ee
where $A$ is the circular area of the cylinder. The charge enclosed by the 
cylinder is easily computed, and one obtains
\be
E(x) =  \left\{\begin{array}{cc} \rho_1 x & \mbox{for}\;\;0<x<L_0 \\[2ex]
(\rho_1-\rho_2)L_0+\rho_2 x & \mbox{for}\;\; L_0<x<L \end{array}\right. \, .
\ee
This yields for the Coulomb energy,
\bea \label{wCslab}
w_C
&=&
\frac{\chi L_0^2}{6}\Big[\rho_1^2(3\chi^{-1}-2)+\rho_2^2(\chi^{-3}-3\chi^{-2}
\non[2ex]
&&+3\chi^{-1}-1)+3\rho_1\rho_2(\chi^{-2}-2\chi^{-1}+1)\Big] \non[2ex]
&=& \frac{(\rho_1-\rho_2)^2L_0^2}{6}(1-\chi)^2 \, ,
\eea
again showing both general and charge neutral expressions. Equations (\ref{wCbubbble}), (\ref{wCrod}), and (\ref{wCslab}) give the result (\ref{WCV}) in the main text. 

\section{Boundary behavior of the profiles}
\label{app:boundary}

In this appendix I derive Eqs.\ (\ref{square}), which show the lowest-order behavior of the profiles at the boundaries of the unit cells. In order to 
compute this behavior, it is useful to write the Euler-Lagrange equations (\ref{dOm}) in the compact vector form
\be \label{dOmvec}
\nabla^2 \vec{v}(x) = \vec{f}[\vec{v}(x)] \, ,
\ee
where $x$ is a Cartesian coordinate for slabs, the cylindrical radial coordinate for rods, or the spherical radial coordinate for bubbles, and 
the Laplacian is
\be \label{laplace}
\nabla^2 = \frac{\partial^2}{\partial x^2} + \frac{d-1}{x}\frac{\partial}{\partial x} \, , 
\ee
where the co-dimension is $d=1,2,3$ for slabs, rods, bubbles, respectively. The components of $\vec{v}$ are the functions $\bar{\sigma},\bar{\omega},\bar{\rho},\mu_e$, and the components of $\vec{f}$ are the right-hand sides of Eqs.\ (\ref{dOm}). Let
$x_0$ be the point at the boundary or the center of the unit cell, and denote  
\be
\vec{v}_0\equiv \vec{v}(x_0) \, , \quad \vec{f}_0\equiv \vec{f}[\vec{v}_0] \, .
\ee
For rods and bubbles one later has to distinguish whether $x_0$ is in the center, $x_0=0$, or at the edge, $x_0>0$, of the unit cell. Until I make this distinction explicitly, the following equations hold for any geometry. A small deviation in the vicinity of $x_0$ is introduced via 
\be
\vec{v}(x) = \vec{v}_0 + \delta \vec{v}(x) \, .
\ee
Then, the linearized version of Eq.\ (\ref{dOmvec}) reads
\be \label{dOmlin}
\nabla^2 \delta \vec{v}(x) = \vec{f}_0+J_f\delta\vec{v}(x) \, ,
\ee
where $J_f$ is the Jacobian matrix of $\vec{f}$ with respect to $\vec{v}$,  
\be
(J_f)_{ij} = \frac{\partial f_i}{\partial v_j} \, , 
\ee
where $i,j = 1, \ldots, 4$. These derivatives can easily be calculated analytically. It will turn out, however, that to lowest order they drop out of the final result and thus there is no need to write them down explicitly. 
The linearized, coupled differential equations (\ref{dOmlin}) can be solved 
by first diagonalizing them, 
\be \label{dOmdiag}
\nabla^2 \delta \tilde{\vec{v}}(x) = \tilde{\vec{f}}_0+U^{-1}J_fU\delta\tilde{\vec{v}}(x) \, ,
\ee
with a $4\times 4$ matrix $U$ which diagonalizes $J_f$, i.e., denoting the eigenvalues of $J_f$ by $\lambda_i$ one has
\be
U^{-1}J_fU = {\rm diag}(\lambda_1,\ldots,\lambda_4) \, , 
\ee
and with 
\be
\delta \tilde{\vec{v}} \equiv U^{-1} \delta \vec{v} \, , \qquad 
\tilde{\vec{f}}_0 \equiv U^{-1}  \vec{f}_0 \, .
\ee
Each component of the vector equation (\ref{dOmdiag}) now reads
\be \label{dropi}
\nabla^2\delta \tilde{v}(x) = \tilde{f}_0+\lambda \delta\tilde{v} (x) \, , 
\ee
where I have omitted the index $i$ at every quantity for notational convenience. In each case, $d=1,2,3$, the solution can be found analytically. One obtains
\be \label{dvtil}
\delta \tilde{v}(x) = -\frac{\tilde{f}_{0}}{\lambda} + C\eta(x) + D \zeta(x) \, ,
\ee
where $C$ and $D$ are integration constants, and  
\begin{subequations} \label{eta}
\bea
\hspace{-1cm}d=1:  && \eta(x) =  e^{\sqrt{\lambda}x} \, , \quad \zeta(x) =e^{-\sqrt{\lambda}x}\, , \\[2ex]
\hspace{-1cm}d=2:  && \eta(x) = I_0(\sqrt{\lambda}x) \, , \quad 
\zeta(x) = K_0(\sqrt{\lambda}x) \, , \label{etarod}\\[2ex]
\hspace{-1cm}d=3:  && \eta(x) =  \frac{e^{\sqrt{\lambda}x}}{x} \, , \quad \zeta(x) = \frac{e^{-\sqrt{\lambda}x}}{x} \, , 
\eea
\end{subequations}
where $I_n$ and $K_n$ are the modified Bessel functions of the first and second kinds, respectively. The integration constants are now determined by the boundary conditions
\be \label{bounddv}
\delta\tilde{v}(x_0)=\delta\tilde{v}'(x_0)=0\,, 
\ee
where prime denotes derivative with respect to $x$.  

At this point, one needs to distinguish between zero and nonzero $x_0$. First assume $x_0>0$, such that all functions and their derivatives are regular at $x_0$. Then, 
the boundary conditions yield 
\be \label{CD}
C=\frac{\tilde{f}_0}{\lambda}\frac{\zeta_0'}{\zeta_0'\eta_0-\eta_0'\zeta_0}\, , 
\quad D=-\frac{\tilde{f}_0}{\lambda}\frac{\eta_0'}{\zeta_0'\eta_0-\eta_0'\zeta_0}
\, , 
\ee
where the subscript 0 denotes evaluation of the function at $x_0$. With the expansions 
\begin{subequations}
\bea
\eta(x) &\simeq& \eta_0+(x-x_0)\eta_0'+\frac{(x-x_0)^2}{2}\eta_0'' \, , \label{etaexp}\\[2ex]
\zeta(x) &\simeq& \zeta_0+(x-x_0)\zeta_0'+\frac{(x-x_0)^2}{2}\zeta_0'' \, , 
\eea
\end{subequations}
one finds 
\bea \label{f0t}
-\frac{\tilde{f}_{0}}{\lambda} + C\eta(x) + D \zeta(x) \simeq \frac{(x-x_0)^2}{2}
\tilde{f}_0 \, , 
\eea
where
\be
\frac{\zeta_0'\eta_0''-\eta_0'\zeta_0''}{\zeta_0'\eta_0-\eta_0'\zeta_0} = \lambda 
\ee
has been used, 
which can be checked explicitly for all three cases with the help of Eqs.\ (\ref{eta}). One now has to remember that Eq.\ (\ref{dropi}) denotes only one of 4 components, and thus, returning to vector notation, the solution of Eq.\ (\ref{dOmdiag}) close to $x_0$ is 
\be
\delta\tilde{\vec{v}}(x) \simeq \frac{(x-x_0)^2}{2}
\tilde{\vec{f}}_0 \,  .
\ee
Multiplying both sides from the left with $U$ and thus undoing the 
rotation that was necessary for the diagonalization, one obtains the simple result 
\bea \label{dv}
\delta\vec{v}(x) \simeq \frac{(x-x_0)^2}{2} \vec{f}_0 \,  .
\eea
Note that only the lowest-order result allows for such a simple expression. The full result (for the linearized problem) is obtained by inserting $C$ and 
$D$ from Eq.\ (\ref{CD}) into $\delta\tilde{v}$ from Eq.\ (\ref{dvtil}), and then undoing the rotation by multiplication with $U$. Since each of the four $\delta\tilde{v}_i$ depends on a different eigenvalue $\lambda_i$, this leads to a linear combination with coefficients which are impossible to write down in a compact way (they can of course be evaluated numerically without problems). Only because the lowest-order result for $\delta\tilde{v}_i$ does {\it not} depend on the eigenvalue $\lambda_i$ it yields such a simple form for the unrotated functions $\delta v_i$.

It remains to discuss the case $x_0=0$. Obviously, in the case of slabs, there is no qualitative difference between the two ends of the domain, i.e., between the center of the slab and the boundary of the unit cell. Therefore, both ends are covered by the result (\ref{dv}). In the center of the rod, where $x_0=0$, the function $\zeta(x)$ from Eq.\ (\ref{etarod}) diverges, and thus regularity requires $D=0$. The boundary conditions (\ref{bounddv}) are fulfilled 
with $C=\tilde{f}_0/\lambda$, and thus (again omitting the index $i$)
\be
\delta\tilde{v}(x) = 
\frac{\tilde{f}_0}{\lambda}\left[I_0(\sqrt{\lambda}x)-1\right] \, .
\ee 
This can easily be expanded and again one is left with a result proportional to $\tilde{f}_0$, such that after reinstating the vector index and undoing the rotation one obtains
\be \label{Crod}
\mbox{rod center:}\qquad \delta\vec{v}(x) \simeq \frac{x^2}{4} \vec{f}_0 \,  .
\ee
Finally, for the center of the bubble, regularity and the boundary conditions (\ref{bounddv}) yield $C=-D=\tilde{f}_0/(2\lambda^{3/2})$, such that
\be
\delta\tilde{v}(x) = \frac{\tilde{f}_0}{\lambda}\left(\frac{\sinh\sqrt{\lambda} x}{\sqrt{\lambda}x}-1\right) \, .
\ee
In analogy to the center of the rod, one thus obtains
\be \label{Cbubble}
\mbox{bubble center:}\qquad \delta\vec{v}(x) \simeq \frac{x^2}{6} \vec{f}_0 \,  .
\ee
The results (\ref{Crod}) and (\ref{Cbubble}) differ from the result 
(\ref{dv}) in the numerical prefactor. It is easy to see that this difference comes from the different form of the Laplacian (\ref{laplace}) and that the number in the numerator is simply $2d$.

In summary, Eqs.\ (\ref{dv}), (\ref{Crod}), and (\ref{Cbubble}) show the behavior of the profiles at the center and the edge of the unit cell for all geometries considered in this paper and are written in the form (\ref{square}) in the main text.  

\bibliography{references}

\end{document}